\documentclass[onefignum,onetabnum]{siamart190516}

\usepackage{lipsum}
\usepackage{amsfonts}
\usepackage{graphicx}
\usepackage{epstopdf}
\usepackage{isomath}
\usepackage{amsmath}
\usepackage{amssymb}
\usepackage{amscd}
\usepackage{amsfonts}

\usepackage{algorithm}
\usepackage{algorithmicx}
\usepackage{algpseudocode}

\usepackage{amsbsy}
\usepackage{stmaryrd}
\usepackage{siunitx}
\usepackage{euscript}
\usepackage[utf8]{inputenc}
\usepackage[T1]{fontenc}
\usepackage{newtxtext} 
\everymath{\displaystyle}
\usepackage{exscale}
\usepackage{array}

\usepackage{subcaption}

\ifpdf
  \DeclareGraphicsExtensions{.eps,.pdf,.png,.jpg}
\else
  \DeclareGraphicsExtensions{.eps}
\fi

\newcommand{\bfdelta}{\mathbold {\delta}}
\newcommand{\bfsigma}{\mathbold {\sigma}}
\newcommand{\bfSigma}{\mathbold {\Sigma}}
\newcommand{\bfepsilon}{\mathbold {\epsilon}}
\newcommand{\bfvarepsilon}{\mathbold {\varepsilon}}

\newcommand{\bfLambda}{\mathbold {\Lambda}}
\newcommand{\bfGamma}{\mathbold {\Gamma}}

\newcommand{\bfmu}{\mathbold {\mu}}
\newcommand{\bfphi}{\mathbold {\phi}}
\newcommand{\bfpsi}{\mathbold {\psi}}
\newcommand{\bfPsi}{\mathbold {\Psi}}

\newcommand{\bfzero}{\mathbf{0}}

\newcommand{\const}{\mathrm{const.}}

\newcommand{\parderiv}[2]{\frac{\partial #1}{\partial #2}}
\newcommand{\dm}{\ \mathrm{d}}

\newcommand{\bfb}{{\mathbold b}}
\newcommand{\bfc}{{\mathbold c}}

\newcommand{\bfg}{{\mathbold g}}

\newcommand{\bfm}{{\mathbold m}}
\newcommand{\bfn}{{\mathbold n}}

\newcommand{\bfp}{{\mathbold p}}

\newcommand{\bfu}{{\mathbold u}}
\newcommand{\bfv}{{\mathbold v}}

\newcommand{\bfx}{{\mathbold x}}
\newcommand{\bfy}{{\mathbold y}}
\newcommand{\bfz}{{\mathbold z}}

\newcommand{\bfA}{{\mathbold A}}
\newcommand{\bfB}{{\mathbold B}}
\newcommand{\bfC}{{\mathbold C}}
\newcommand{\bfD}{{\mathbold D}}

\newcommand{\bfG}{{\mathbold G}}
\newcommand{\bfH}{{\mathbold H}}
\newcommand{\bfI}{{\mathbold I}}

\newcommand{\bfK}{{\mathbold K}}
\newcommand{\bfL}{{\mathbold L}}
\newcommand{\bfM}{{\mathbold M}}
\newcommand{\bfN}{{\mathbold N}}

\newcommand{\bfQ}{{\mathbold Q}}

\newcommand{\bfV}{{\mathbold V}}

\DeclareMathOperator{\divergence}{div}

\DeclareMathOperator{\trace}{tr}
\DeclareMathOperator*{\argmin}{argmin}

\DeclareMathOperator{\E}{\mathbb{E}}

\newsiamremark{remark}{Remark}
\newsiamremark{hypothesis}{Hypothesis}
\crefname{hypothesis}{Hypothesis}{Hypotheses}
\newsiamthm{claim}{Claim}

\headers{High-dimensional Nonlinear Bayesian Inference}{M. Karimi, M. Massoudi, K. Dayal, and M. Pozzi}

\title{High-dimensional Nonlinear Bayesian Inference of Poroelastic Fields from Pressure Data}

\author{
Mina Karimi\thanks{Department of Civil and Environmental Engineering, Carnegie Mellon University 
(\email{minakari@andrew.cmu.edu}).}
\and Mehrdad Massoudi\thanks{National Energy Technology Laboratory, Pittsburgh PA 15236-0940.}
\and Kaushik Dayal\thanks{Department of Civil and Environmental Engineering, Carnegie Mellon University; Center for Nonlinear Analysis, Department of Mathematical Sciences, Carnegie Mellon University; Department of Mechanical Engineering, Carnegie Mellon University; Scott Institute for Energy Innovation, Carnegie Mellon University.}
\and Matteo Pozzi\thanks{Department of Civil and Environmental Engineering, Carnegie Mellon University; Scott Institute for Energy Innovation, Carnegie Mellon University.}
}

\usepackage{amsopn}

\ifpdf
\hypersetup{
  pdftitle={High-dimensional Nonlinear Bayesian Inference of Poroelastic Fields
from Pressure Data},
  pdfauthor={M. Karimi, M. Massoudi, K. Dayal, and M. Pozzi}
}
\fi

\begin{document}

\maketitle

\begin{center}
    \textbf{To appear in Mathematics and Mechanics of Solids (\url{https://doi.org/10.1177/10812865221140840})}    
\end{center}

\begin{abstract}
    We investigate solution methods for large-scale inverse problems governed by partial differential equations (PDEs) via Bayesian inference. The Bayesian framework provides a statistical setting to infer uncertain parameters from noisy measurements. To quantify posterior uncertainty, we adopt Markov Chain Monte Carlo (MCMC) approaches for generating samples.
    To increase the efficiency of these approaches in high-dimension, we make use of local information about gradient and Hessian of the target potential, also via Hamiltonian Monte Carlo (HMC). Our target application is inferring the field of soil permeability processing observations of pore pressure, using a nonlinear PDE poromechanics model for predicting pressure from permeability.
    We compare the performance of different sampling approaches in this and other settings. We also investigate the effect of dimensionality and non-gaussianity of distributions on the performance of different sampling methods.  
\end{abstract}


\section{Introduction and Background}

Many problems in science and engineering problems can be modeled by partial differential equations (PDEs). Inverse problems constrained by PDEs are a challenging class of these problems, which play an essential role in investigating many physical systems, including geomechanical engineering, medical engineering, astrophysics, e.g. ~\cite{petra2011model, martin2012stochastic, bui2013computational, petra2014computational, alghamdi2020bayesian, villa2019hippylib, funke2013framework, cui2014likelihood, jha2022mutual, jha2020bayesian,jha2022goal}.
Inverse problems aim at inferring the unknown model parameters of a physical system from observations, which may be limited, indirectly related to the parameters, and affected by noise.
When the parameter domain is high-dimensional, and the relation between parameters and observations is defined by a complex mechanics model, 
solving inverse problems is computationally expensive.
One approach for addressing these problems is to identify a point-based estimator via the minimizing of a function quantifying the discrepancies between model predictions and observations, e.g. the negative log-likelihood function.
However, in this formulation the solution may not be unique, and the point-based estimate may be not informative on the posterior knowledge.
One classical method to overcome the ill-posedness and to decrease the sensitivity of the solution to measurements' noise is to add a regularization term, enforcing continuity and smoothness of the solution  ~\cite{petra2011model, engl1996regularization, tarantola2005inverse}.
In the Bayesian framework, the unknown parameters are treated as random variables, with an assigned prior probability, while the likelihood function relies on the forward model and the noisy measurements ~\cite{martin2012stochastic, villa2019hippylib}.
Prior and likelihood are integrated in the posterior distribution.
A point-based estimator for the posterior evaluation is the Maximum a Posteriori (MAP), but the posterior uncertainty can also be represented, describing the posterior distribution via samples or approximations, when it cannot be derived in exact form.

In this paper, we adopt a Bayesian inference framework to infer the uncertain parameters of a PDE-based forward poromechanics model.
For solving the Bayesian inverse problem numerically, the infinite-dimensional parameter space has to be discretized to a finite-dimensional domain.
Different discretization methods have been studied in the context of infinite-dimensional inverse problems.
Finite element discretization methods use a finite number of continuous Lagrange basis functions to approximate the infinite-dimensional parameter space ~\cite{bui2013computational, villa2019hippylib}.
The Karhunen-Loeve (K-L) expansion provides an alternative approach, representing the infinite-dimensional parameter set in terms of its eigenvalues, and allowing of truncating after a finite number of terms and approximate dimensionality reduction ~\cite{uribe2020bayesian, uribe2020bayesian1}.  

We discuss the effectiveness of sampling methods based on Markov Chain Monte Carlo (MCMC) for representing the posterior distribution in the high-dimensional parameter space.
The standard Metropolis-Hastings algorithm is computationally too expensive for our  high-dimensional application ~\cite{gebraad2020bayesian}.
There have been many efforts to accelerate the speed of sampling for high-dimensional inverse problems, such as developing reduced-order models ~\cite{arridge2006approximation, bui2008parametric, galbally2010non, patsialis2020bayesian}, using local gradient information as Hamiltonian Monte Carlo (HMC) method ~\cite{betancourt2017conceptual}, Hessian information as Riemannian manifold HMC ~\cite{lee2018convergence, bui2014solving}, and Hessian based MCMC methods ~\cite{qi2002hessian, martin2012stochastic, bui2013computational, petra2014computational}.
In this paper, we investigate the efficiency of Hessian-based MCMC and Hessian-based HMC methods in high dimensional problems. 

Using Hessian information can significantly improve the performance of MCMC methods in infinite-dimensional inverse problems.
Qi and Minka ~\cite{qi2002hessian} proposed a Hessian based Metropolis-Hasting (HMH) algorithm.
They formulated an adaptive proposal density by approximating a Gaussian distribution using the local gradient and Hessian information.
Martin et al. ~\cite{martin2012stochastic} applied the HMH method for high-dimensional inverse problems by approximating the low-rank local Hessian to ensure the positive definiteness of the matrix, obtaining what they call the Stochastic Newton MCMC (SNMCMC) method.
Petra et al. ~\cite{petra2014computational} modified the SNMCMC method by using the Hessian at the MAP point to decrease its computational cost. 

The HMC method can explore the posterior distribution faster than regular MCMC alternatives, as it uses local gradient information to make long-distance moves through the parameter space ~\cite{betancourt2017conceptual}.
HMC algorithm can be used for efficient sampling in high-dimensional parameter spaces by selecting appropriate; step size, total number of samples, and mass matrix ~\cite{gebraad2020bayesian}.
Several works have proposed to modify the HMC algorithm by tuning the mass matrix using Hessian information ~\cite{fichtner2018tutorial, gebraad2020bayesian, lee2018convergence, wang2013adaptive, zhang2011quasi}.
Lee and Vempala ~\cite{lee2018convergence} proved that the Riemannian Manifold HMC (RMHMC) accelerates the convergence rate of HMC by using the local Hessian information as the mass matrix, and Zhang and Sutton ~\cite{zhang2011quasi} developed a modified HMC method by using the approximated BFGS Hessian as the mass matrix. Bui-Thanh and Girolami ~\cite{bui2014solving} investigate a RMHMC method using the exact and low-rank Fisher information matrix.

We particularly study the performance of Hessian-informed MCMC and HMC methods that have been proposed recently for high-dimensional inverse problems. We compare the SN-MAP method presented by Petra et al. ~\cite{petra2014computational}, the Metropolis- Adjusted Langevin algorithm (MALA) ~\cite{roberts1998optimal} algorithm using the Hessian information at the MAP point, and the H-HMC method presented by Bui-Thanh and Girolami ~\cite{bui2014solving}. We investigate several numerical examples to show how the dimension and non-Gaussian nature of the problem can affect the performance of different Hessian-informed methods. We avoid investigating Hessian-informed methods using the local Hessian information since those are computationally expensive and require calculating the second derivative information in each iteration.

There are many applications for high-dimensional inverse problems ~\cite{alghamdi2020bayesian, bui2014solving, martin2012stochastic, petra2014computational}. Our analysis targets the inference of soil permeability in regions close to injection sites related to a broad range of energy activities such as; waste-water injection, CO2 sequestration, geothermal energy activities, hydraulic fracking.
Inferring these parameters is crucial for estimating the capacity of reservoirs, predicting the pore pressure and stress around the injection centers, and assessing the hazards of sliding and injection-induced seismic events ~\cite{sumy2017low, schoenball2017systematic, keranen2018induced, segall2015injection, wang2015statistical, wang2016bayesian, langenbruch2018physics, keranen2013potentially, li2006co2, shapiro2015fluid, broccardo2020induced}.
We infer the unknown permeability field from the collected pressure measurements.
In our work, we adopt the Bayesian framework to infer the unknown poroelastic properties of a nonlinear poromechanics model from the noisy and sparse measurements.
The MAP is identified using a inexact Newton solver, then samples are generating via MCMC approaches, starting from that point.
We apply the developed model to solve a large-scale nonlinear inverse problem and determine the unknown properties of the deep underground layers. 

\paragraph*{Organization} Section \ref{sec:Bayesian framework} provides the general Bayesian formulation for high-dimensional inverse problems, and discusses the choice of prior. Section \ref{sec:Sampling methods} reviews and discusses the standard and accelerated sampling methods for exploring the posterior distribution. Section \ref{sec:Forward model} provides the governing equations of the forward poromechanics model. Section \ref{sec:Numerical Results} illustrates the numerical results of identifying the MAP point and generating samples from prior and posterior distributions.

\section{Bayesian Framework for High-Dimensional Inverse Problems}\label{sec:Bayesian framework}
We infer field $\theta(x)$, where $x$ is a spatial coordinate on reference domain $\Omega$, following the Bayesian framework.
The prior knowledge about the field is modeled by a function $\pi(\theta)=\text{log}\big(p(\theta)\big)$, and $p(\theta)$ is assumed to be Gaussian, so that:
\begin{equation}\label{eqn:general prior function}
    \pi(\theta) = -\frac{1}{2}||\mathcal{A}(\theta-m_\pi) ||^2  + \const
\end{equation}
where $m_\pi(x)$ is the prior mean function and $\mathcal{A}$ is a Laplacian-like operator, that will be defined in Section \ref{sec:choice of prior}. 
For any function $f \in L^2(\Omega)$, $L^2(\Omega)$ being the space of square integrable functions, the norm is defined as:
\begin{equation}
    ||f(x)||^2 = \int_\Omega f^2(x) \dm x
\end{equation}

A dataset of $n$ noisy observations is available, listed in a vector $\bfy=[y_1,y_2,\cdots,y_n]^\top$.
The observation $i$ is related to the field via equation $y_i = Q_\theta (x_i, t_i) +\epsilon_i$,
where the function $Q_\theta$ predicts the value to be measured, as a function of the parameter field $\theta$, time $t$ 
and location $x$, and it is computed by solving the forward predictive model. Noise terms are listed in a vector $\bfepsilon=[\epsilon_1,\epsilon_2,\cdots,\epsilon_n]^\top$, modeled as a Gaussian random variable, independent of $\theta$, with a zero mean and covariance matrix $\bfGamma_{\epsilon}$, i.e., $\bfepsilon \sim \mathcal{N} (\bfzero,\bfGamma_{\epsilon})$.

Probabilistically, the agreement between predictions and measurements is modeled by the log-likelihood function $l(\theta)=\text{log}\big(p(\bfy|\theta)\big)$:
\begin{equation}
\label{eq:LH}
    l(\theta) = -\frac{1}{2}||\bfGamma^{-1/2}_\epsilon(\bfQ_\theta-\bfy)||^2  + \const
\end{equation}
where $\bfQ_\theta = [Q_\theta(x_1,t_1),\cdots,Q_\theta(x_n,t_n)]^\top$ is the vector of predictions. In this paper, for each positive definite and symmetric matrix $\bfG$, the matrix $\bfG^{-1/2}$ is a decomposition of inverse of that matrix, so that $\bfG^{-1}=\bfG^{-1/2} \bfG^{-\top/2}$ (e.g. the decomposition can be based on eigenvalue analysis or on Cholesky's method). 
For column vectors, the norm is defined as $||\bfv||^2 = \bfv^\top \bfv$.
If $\bfGamma_\epsilon=\sigma_\epsilon^2\bfI$ (where $\bfI$ is the identity matrix), then $\bfGamma^{-1/2}_\epsilon$ can be replaced by $\sigma^{-1}_\epsilon$ in Eq. \eqref{eq:LH}.

Following the Bayes' rule, the log-posterior density $\omega(\theta)=\text{log}\big(p(\theta|\bfy)\big)$ is:
\begin{equation}
    \omega(\theta) = l (\theta) + \pi(\theta) +  \text{const.}
\end{equation}
Motivated by the equation above, we define the objective function as $\mathcal{J}(\theta)= - \omega(\theta) + \text{const.}$:
\begin{equation}\label{eqn:general objective function}
    \mathcal{J} (\theta) = \frac{1}{2}||\bfGamma^{-1/2}_\epsilon(\bfQ_\theta-\bfy)||^2+\frac{1}{2}||\mathcal{A}(\theta-m_\pi) ||^2
\end{equation}
The MAP point is $\theta_\text{MAP}=\argmin \big(\mathcal{J}(\theta)\big)$, and it is a parameter field that maximizes the posterior density.
To identify $\theta_\text{MAP}$, we minimize $\mathcal{J}$: while least squares approaches minimize the first term (based on the likelihood), the second one (based on prior knowledge) usually acts as a regularization.

\subsection{Choice of Prior Model}\label{sec:choice of prior}
The choice of $p(\theta)$ is crucial as it allows to include the prior knowledge on the parameter field $\theta$ in a Bayesian inference setting.
In this paper, we use a continuous Gaussian measure $\mathcal{N} (m_\pi, \mathcal{C}_\pi)$
instead of adopting a discretized Gaussian distribution and covariance matrix ~\cite{bui2013computational, bui2014solving}. 
Here, $\mathcal{C}_\pi$ is the covariance operator which is defined as below:
\begin{equation}
    (\mathcal{C}_\pi \phi) (\bfx) = \int_{\Omega} \bfc(\bfx,\bfy) \bfphi(\bfy) \dm \bfy
\end{equation}
where $\bfc(\bfx,\bfz)$ is covariance function:
\begin{equation}
    \bfc(\bfx,\bfz) = \E \big[ \big( \theta (\bfx) - m_\pi (\bfx) \big) \big( \theta (\bfz) - m_\pi (\bfz) \big)  \big]
\end{equation}
Given measure function $\mu_\pi$, the operator is defined as:
\begin{equation}\label{eqn:covarian operator}
    \mathcal{C}_\pi = \int_\Omega (\theta - m_\pi) \otimes (\theta- m_\pi) \mu_\pi(\dm\theta) = \E [(\theta - m_\pi) \otimes (\theta- m_\pi)]
\end{equation}

The operator $\mathcal{C}_\pi$ is fast-to-apply and allows a simple discretization in high dimension  ~\cite{bui2013computational}. As it is described in Appendix \ref{sec:Appendix bA} the prior function $\pi$ (Eq. \eqref{eqn:general prior function}) is written as follows: 
\begin{equation}
    \pi(\theta) = - \frac{1}{2}|| \mathcal{C}_\pi^{-1/2} (\theta - m_\pi) ||^2 + \text{const.}
\end{equation}
Commonly, the covariance operator is considered as the fractional power of Laplacian-like operator $\mathcal{A}$, i.e. $\mathcal{C}_\pi = \mathcal{A}^{-\alpha}~, ~\alpha>0$, where $\mathcal{A}$ is positive definite, self adjoint and invertible ~\cite{dashti2013bayesian, pandit2019comparative, cressie2015statistics}.
Here we choose the covariance operator as $\mathcal{C}_\pi =\mathcal{A}^{-2}$, so that $\mathcal{C}_\pi^{-1/2} =\mathcal{A}$, and the action of operator $\mathcal{A}$ on $\theta$, i.e., $\mathcal{A}\theta$ is defined by the following elliptic Boundary Value Problem ($\mathcal{A} = -\gamma {\bfPsi} \Delta +\delta\bfI$):
\begin{align}\label{eqn: 2.10}
    -\gamma \divergence (\bfPsi \nabla \theta ) + \delta \theta = s ~~ \text{in}~ \Omega\\\nonumber
    \gamma (\nabla \theta)\cdot\bfn  = 0 ~~ \text{on} ~\partial\Omega
\end{align}
where $\Delta$ is the Laplacian operator, $\gamma$ and $\delta$ are positive (hyper)parameters, $s$ is a white noise field and the random $\theta$ field can be expressed as linear function of the random noise, as $\mathcal{A}^{-1} s$.
Here, $\bfn$ is the unit normal vector on $\partial \Omega$. The symmetric and positive-definite second order tensor $\bfPsi$ defines the anisotropy, and indicates the direction $\hat{\bfm}$ in which the parameter space $\theta$ has a larger variation. 
$\bfPsi$ can then be defined in terms of the unit vector $\hat{\bfm}$ as follows:
\begin{equation}
    \bfPsi = a\left( \bfI - (1-b/a)\hat{\bfm} \otimes \hat{\bfm}^\top \right)
\end{equation}
In a two-dimensional setting, $\bfPsi$ is can be represented as:
\begin{equation}
    \bfPsi = \begin{bmatrix}
              a\sin{(\beta)}^2 + b\cos{(\beta)}^2 & (b-a)\sin{(\beta)}\cos{(\beta)} \\
              (b-a)\sin{(\beta)}\cos{(\beta)} & b\sin{(\beta)}^2 + a\cos{(\beta)}^2 
             \end{bmatrix}
\end{equation}
where $a$, $b$ and $\beta$ are constants.
The larger $\gamma$ and $\delta$ are, the smaller becomes the variance (Eq. (\ref{eqn: 2.10})). 
Based on the above formulation, the continuous form of prior term can be written as:
\begin{equation}\label{eqn:operator A}
    \mathcal{A} \theta = \begin{cases} -\gamma \divergence (\bfPsi\nabla\theta) +\delta\theta & \text{ in } \Omega\\ \gamma(\nabla\theta)\cdot\bfn  & \text{ on } \partial\Omega \end{cases}
\end{equation}

\subsection{Discretization of the field}
We use a finite element discretization with continuous Lagrangian basis functions $\{\phi_j\}_{j=1}^m$, and the corresponding nodal values $\{\bfx_j\}_{j=1}^m$, to obtain the finite-dimensional approximation of the problem.
Thus, the discretized field can be described as a vector of coefficients ${\bfpsi} = [\psi_1, \dots, \psi_m]^\top$.
The inner product between nodal coefficient vectors can be weighted by a positive-definite and symmetric matrix $\bfN$, with entries given by:
\begin{equation}
    N_{ij} = \int_\Omega\phi_i(\bfx)\phi_j(\bfx)\dm \bfx
\end{equation}
Then, for every pair of functions $\theta_1, \theta_2 \in L^2(\Omega)$, we can define the weighted inner product in the discretized form as $(\theta_1,\theta_2)_{L^2(\Omega)} \approx (\bfpsi_1,\bfpsi_2)_\bfN = \bfpsi_1^\top\bfN\bfpsi_2$.
The matrix representation of the the Laplacian operator $\mathcal{A}$ can be defined as $\bfA = \bfN^{-1} \bfK$ ~\cite{petra2014computational}, with entry $\{i,j\}$ of $\bfK$ as:
\begin{equation}\label{eqn:matrix of operator A}
    K_{ij} = \int_\Omega \big[ \gamma \bfPsi \nabla\phi_i (\bfx)\cdot \nabla\phi_j (\bfx) + \delta \phi_i(\bfx)\phi_j(\bfx) \big]\dm\bfx
\end{equation}
Therefore, the log prior term can be approximated as:
\begin{equation}
    \pi(\theta) \cong - \frac{1}{2}||\bfA^{-1/2} ~ \bfpsi||^2 + \text{const.}
\end{equation}
and the objective function can be expressed in terms of $\bfpsi$ as:
\begin{equation}\label{eqn:objective function discrete}
    \mathcal{J} (\bfpsi) = \frac{1}{2}||\bfGamma^{-1/2}_\epsilon(\bfQ_\bfpsi-\bfy)||^2 + \frac{1}{2}||\bfA^{-1/2} ~ \bfpsi||^2
\end{equation}
where $\bfQ_\bfpsi$ is the predictive vector, reconstructing field $\theta=\sum^m_{j=1}\psi_j\phi_j$.
To minimize this nonlinear function with respect to $\theta$, quasi-Newton methods can be generally used. 

\section{Sampling Methods for High-dimension Inverse Problems}\label{sec:Sampling methods}
In this section we review some key methods for drawing samples from probability density functions.
We discuss the Metropolis-Hastings MCMC and Hessian based MCMC methods, and compare their computational costs and efficiency for sampling high-dimensional parameter space.

\subsection{Metropolis Hastings Method}
The Metropolis Hastings MCMC (MH-MCMC) method (Algorithm \ref{alg:MHMCMC}) relies on proposal density $q$, which samples next value of random variables $\bfz$, along a chain, centered at current location $\bfc$.
One of the most popular choices of proposal (log) density is the isotropic Gaussian distribution:
\begin{equation}
   q(\bfc,\bfz) = - \frac{1}{2} ||\Delta t^{-1} (\bfc - \bfz) ||^2 + \text{const.}
\end{equation}
where $\Delta t$ is a fixed step-size. 

\begin{algorithm}[h]
\caption{Metropolis-Hastings MCMC}\label{alg:MHMCMC}

\begin{algorithmic}[1]
\Require Initial parameter $\bfpsi_0$
\State Compute $\mathcal{J}_0=\mathcal{J}(\bfpsi_0)$,
\For {$k=0, \cdots, N$}
\State Draw sample $\bfz$ from proposal density $q(\bfpsi_k, \cdot)$
\State Compute $\mathcal{J}_z=\mathcal{J}(\bfz)$, $\Delta q=q(\bfz,\bfpsi_k)-q(\bfpsi_k,\bfz)$,
\State Compute $\alpha = -\mathcal{J}_z+\mathcal{J}_k+\Delta q$
\State Compute $a = \min [1,\exp{ \left( \alpha \right)}]$
\State Draw $u$ from $\mathcal{U}([0, 1))$ 
\If {$u < a$}
    \State Accept: Set $\bfpsi_{k+1} = \bfz$, $\mathcal{J}_{k+1} = \mathcal{J}_z$
\Else
    \State Reject: Set $\bfpsi_{k+1} = \bfpsi_k$, $\mathcal{J}_{k+1} = \mathcal{J}_k$
\EndIf
\EndFor
\Ensure Random walk: $\{\bfpsi_k\}_{k=1}^N$
\end{algorithmic}
\end{algorithm}
In high-dimensional problems governed by PDEs, the posterior standard deviation of a linear combination of the field variables varies significantly depending of the direction of the combination, and this poses some challenges in choosing the step size.
Selecting a large step size will induce a low acceptance rate, due to the small standard deviation of some directions, and a step size as small as the smallest standard deviation will induce a slow mixing (and, since we need to solve the forward model in each iteration, the simulation campaign will be computationally expensive).

In the following sections, we discuss some accelerated sampling methods using the local gradient and Hessian information of the objective function $\mathcal{J}$.

\subsection{ Hessian-based MCMC Methods}
In this section we review three Hessian informed MCMC methods proposed by Ghattas and coworkers ~\cite{martin2012stochastic, bui2013computational, petra2014computational}:

The local quadratic form of Eq. \eqref{eqn:general objective function}, based on the Newton method at the vicinity of a current sample $\bfpsi_k$ can be approximated using the Laplace approximation (based on a second order Taylor series) as follows ~\cite{murphy2012machine}:
\begin{equation}\label{eqn: 3.2}
    \mathcal{J}(\bfpsi)\approx\Tilde{\mathcal{J}}(\bfpsi) = \frac{1}{2}|| \bfH^{1/2}_k(\bfpsi-\bfpsi_k)||^2 + \bfg^\top_k (\bfpsi-\bfpsi_k) + \mathcal{J}(\bfpsi_k)
\end{equation}
where $\bfg_k = \nabla \mathcal{J}(\bfpsi_k)$ and $\bfH_k = \nabla^2 \mathcal{J}(\bfpsi_k)$ are the gradient and the Hessian matrix of the objective function, respectively, and $\Tilde{\mathcal{J}}$ is the approximate function.
We note that the Hessian matrix is positive definite at the MAP point. However, we cannot guarantee a positive definite matrix at any arbitrary point.
Therefore, to guarantee a positive definite Hessian matrix, we can use a low-rank approximation  $\Tilde{\bfH}_k$ (Appendix \ref{sec:Appendix A}), which is calculated based on the eigenvalue decomposition, truncating negative and zero eigenvalues.
By expanding Eq. (\ref{eqn: 3.2}), and adding and subtracting the term $\frac{1}{2}\bfg^T_k \Tilde{\bfH}^{-1}_k \bfg_k$, we obtain:
\begin{equation}\label{eqn: 3.3}
    \Tilde{\mathcal{J}}(\bfpsi) = \frac{1}{2}(\bfpsi-\bfpsi_k)^\top \Tilde{\bfH}_k (\bfpsi-\bfpsi_k) + \bfg^\top_k (\bfpsi-\bfpsi_k) + \mathcal{J}(\bfpsi_k) + \frac{1}{2}\bfg^T_k \Tilde{\bfH}^{-1}_k \bfg_k - \frac{1}{2}\bfg^T_k \Tilde{\bfH}^{-1}_k \bfg_k
\end{equation}
Therefore, Eq. (\ref{eqn: 3.3}) can be re-written as ~\cite{martin2012stochastic}:
\begin{equation}
    \Tilde{\mathcal{J}}(\bfpsi) = \frac{1}{2}|| \Tilde{\bfH}^{1/2}_k~(\bfpsi-\bfpsi_k+\Tilde{\bfH}^{-1}_k\bfg_k)||^2 + \const
\end{equation}
where term $\Tilde{\bfH}^{-1}_k\bfg_k$ can be interpreted as the Newton step.
Inspired by this approximation, the Stochastic Newton MCMC (SN-MCMC) method adopts the following proposal log density:
\begin{equation}\label{eqn:modified density MCMC}
    q(\bfpsi_k,\bfz) = - \frac{1}{2}|| \Tilde{\bfH}^{1/2}_k ~ (\bfz - \bfpsi_k+\Tilde{\bfH}_k^{-1}\bfg_k) ||^2 +\const
\end{equation}
When implementing the method, we need to solve the forward problem at each iteration of the sampling process, and calculate the local Hessian, local gradient, and approximated Hessian at each iteration.

Since recalculation of Hessian and approximated Hessian at each sample point is expensive, Bui et al. ~\cite{bui2013computational, petra2014computational} proposed a modified method which is based on always using the Hessian at the MAP point.
Therefore, in the Stochastic Newton MCMC with MAP-based Hessian (SN-MAP) method we rewrite the proposal log density function as follows:
\begin{equation}\label{eqn:proposal SN-MAP}
    q(\bfpsi_k,\bfz) = - \frac{1}{2}|| {\bfH}^{1/2}_\text{MAP} ~ (\bfz - \bfpsi_k+\bfH_\text{MAP}^{-1}~\bfg_k) ||^2 + \const
\end{equation}
Thus, implementing this approach at each iteration we need to run the forward problem, and compute the local gradient at each sampled point, but we calculate the Hessian matrix just once, at the MAP point.

Using the locally approximated proposal density can increase the convergence speed, respect to MH-MCMC. However, when the target distribution is far from a Gaussian distribution, changing the mean value of the proposal density by adding term $(\bfH_\text{MAP}^{-1}\bfg_k)$, see Eq. (\ref{eqn: 3.2}), can decrease the speed of convergence or the sampling process can be stuck in a low probability region.
In order to overcome these problems, we can define a small random learning rate $\gamma \leq 1$, substituting term $(\bfH_\text{MAP}^{-1}\bfg_k)$ with term $(\gamma\bfH_\text{MAP}^{-1}\bfg_k)$, or use a trust region method ~\cite{qi2002hessian}. 

Another Hessian informed MCMC method proposed by Bui et al. ~\cite{bui2013computational, petra2014computational}, called Independence Sampling with MAP point-based Gaussian proposal
(IS-MAP), neglects recalculating the local gradient at each sample by adopting a proposal density centered at the MAP point:
\begin{equation}
    q(\bfpsi_\text{MAP},\bfz) = - \frac{1}{2}|| \bfH^{1/2}_\text{MAP} ~ (\bfz - \bfpsi_\text{MAP}) ||^2 + \const
\end{equation}
where $\bfpsi_\text{MAP}$ is the parameter value at the MAP point.
Therefore, by avoiding recalculation of local Hessian and gradient at each sampling point, we need to calculate the Hessian at the MAP point only once and solve forward problem at each sampling point. 
Note that, in this method, samples are generated independently at each step, so the sample generation is simpler than that of any other MCMC method. However, the acceptance criteria is still the traditional one, for MH-MCMC. This method can be related to the classical rejection sampling approach. 

In the Metropolis-Adjusted Langevin Algorithm (MALA) method ~\cite{girolami2011riemann, roberts1998optimal, roberts1996exponential}, the proposal density is defined as:
\begin{equation}\label{eqn:proposal MALA}
    q(\bfpsi_k,\bfz) = - \frac{1}{2}|| (2 \tau \bfB)^{-1/2} ~ (\bfz - \bfpsi_k + \tau\bfB \bfg_k) ||^2 + \const
\end{equation}
where $\bfB$ is a positive-definite preconditioning matrix, and $\tau$ is a fixed time step, which has to be taken as $0<\tau\leq 1$. 
If we use the inverse of local Hessian information at the MAP as the preconditioning matrix, the proposal density becomes:
\begin{equation}\label{eqn:proposal MALA MAP}
    q(\bfpsi_k,\bfz) = - \frac{1}{2}|| \big(\bfH_\text{MAP}/(2\tau)\big)^{1/2} ~ (\bfz - \bfpsi_k + \tau\bfH_\text{MAP}^{-1}\bfg_k) ||^2 + \const
\end{equation}
The scalar parameter $\tau$ in Eqs. (\ref{eqn:proposal MALA}), and (\ref{eqn:proposal MALA MAP}) can play a similar role as the small learning rate $\gamma$, by preventing the algorithm from being stuck in the low probability regions and yield better results for non-Gaussian distributions.

\subsection{Hamiltonian Monte Carlo Method}
HMC is one of the efficient sampling methods relying on local gradient information.
In the HMC algorithm (Algorithm \ref{alg:HMC}), we define an auxiliary momentum value, then we update the position and momentum according to the Hamiltonian's system of differential equations.
For a general system, the position is the variable of interest (e.g. the parameters to be inferred). The potential energy is defined as the objective function of Section \ref{sec:Bayesian framework}, as the negative logarithm of the target distribution ~\cite{gelman2013bayesian, neal2011mcmc}.
The Hamiltonian function can be defined as:
\begin{equation}\label{eqn:Hamiltonian}
    \mathcal{H}(\bfpsi,\bfp) = \mathcal{K}(\bfpsi,\bfp) + \mathcal{J}(\bfpsi)
\end{equation}
where $\mathcal{K}(\bfpsi,\bfp)$ is the kinetic energy and $\bfp$ is the momentum vector. In order to update variables $\bfp$ and $\bfpsi$, we need to calculate the derivatives of these variables with respect to the "pseudo time" variable $t$.
Therefore, Hamiltonian equations are:
\begin{eqnarray}
   &\frac{d\bfpsi}{dt} = \frac{\partial \mathcal{H}}{\partial\bfp} = \frac{\partial \mathcal{K}}{\partial \bfp} + \frac{\partial \mathcal{J}}{\partial \bfp} = \frac{\partial \mathcal{K}}{\partial \bfp}\\
   &\frac{d\bfp}{dt} = -\frac{\partial \mathcal{H}}{\partial\bfpsi} = -\frac{\partial \mathcal{K}}{\partial \bfpsi} - \frac{\partial \mathcal{J}}{\partial \bfpsi} = - \frac{\partial \mathcal{J}}{\partial \bfpsi}
\end{eqnarray}
Different choices of $\mathcal{K}$ function have been proposed in the literature. One of the most common choices is the \textit{Euclidean-Gaussian} kinetic energy ~\cite{betancourt2017conceptual}:
\begin{equation}
   \mathcal{K}(\bfpsi,\bfp) = \frac{1}{2} ||\bfM^{-1/2} \bfp||^2 + \frac{1}{2}\log |\bfM| + \const
\end{equation}
where $\bfM$ is a positive-definite matrix known as the "mass matrix".
The choice of the mass matrix influences the convergence speed and the performance of the sampling process. 

\subsubsection{Hessian-based HMC Methods}
It has been proved that we can reach a better convergence speed in linear problems by setting the mass matrix as the inverse of posterior covariance  ~\cite{fichtner2018tutorial}.
In a more general setting, Riemannian manifold HMC method uses a \textit{Riemannian-Gaussian} kinetic energy ~\cite{betancourt2013general} to choose the mass matrix as $\bfM = \bfC(\bfpsi)$,
where $\bfC(\bfpsi)$ is a function of the parameter of interest which has to be updated with the local information ~\cite{betancourt2017conceptual}. 

When we consider $\bfC(\bfpsi) = \nabla^2 \mathcal{J}(\bfpsi)$, the Hamiltonian updating equations will induce an updating step similar to that of Newton method ~\cite{lee2018convergence}, which can significantly improve the sampling performance.
Using the Riemannian manifold HMC for high-dimensional inverse problems is computationally expensive since we need to calculate the local Hessian of the posterior at each iteration.

To decrease the computational cost, we can use the Hessian of the posterior at the MAP point as a mass matrix ~\cite{bui2014solving}, which prevent calculating Hessian in each iteration.

\begin{algorithm}[h]

\caption{Hamiltonian Monte Carlo}\label{alg:HMC}

\begin{algorithmic}[1]
\Require Initial parameter $\bfpsi_0$
\For {$k=0, \cdots, N$}
\State Draw momentum sample $\bfp_k$ from proposal density $\mathcal{N}(\bfzero,\bfM)$
\State Compute $\mathcal{H}(\bfpsi_k,\bfp_k)$, $\parderiv{\mathcal{J}}{\bfpsi_k}$, $\parderiv{\mathcal{J}}{\bfpsi_{k+1}}$
\State Update $\bfpsi$ (based on leapfrog algorithm)
\State$\bfp_{k+\frac{1}{2}} =\bfp_k - \frac{\Delta t}{2}\parderiv{\mathcal{J}}{\bfpsi_k}$ 
\State $\bfpsi_{k+1} =\bfpsi_k + \Delta t \bfM^{-1}\bfp_{k+\frac{1}{2}}$
\State $\bfp_{k+1} = \bfp_{k+\frac{1}{2}}-\frac{\Delta t}{2}\parderiv{\mathcal{J}}{\bfpsi_{k+1}}$
\State Compute $\mathcal{H}(\bfpsi_{k+1},\bfp_{k+1})$
\State Compute $\alpha = \mathcal{H} (\bfpsi_k,\bfp_k) - \mathcal{H}({\bfpsi}_{k+1},\bfp_{k+1})$
\State Compute $a = \min \left[1, \exp{ \left( \alpha \right)}\right]$
\State Draw $u$ from $\mathcal{U}([0, 1))$ 
\If {$u < a$}
    \State Accept: $\bfpsi_{k+1}$
\Else
    \State Reject: Set $\bfpsi_{k+1} = \bfpsi_k$
\EndIf
\EndFor
\end{algorithmic}

\end{algorithm}

\subsection{Comparison of SN-MCMC, MALA, and H-HMC Sampling Methods}
This section compares the formulation of MCMC sampling methods relying on Hessian information for high-dimension parameter spaces.
We discuss their similarities and differences.
\subsubsection{SN-MAP}
Following Eq. \eqref{eqn:proposal SN-MAP}, the updating step of SN-MAP method is (here we directly indicate the next candidate point as ${\bfpsi}_{k+1}$, instead of as $\bfz$):
\begin{equation}\label{eqn:up SN-MAP}
    {\bfpsi}_{k+1} ={\bfpsi}_k - {\bfH}_\text{MAP}^{-1} \bfg_k + {\bfH}_\text{MAP}^{-1/2} \bfu  
\end{equation}
where $\bfu$ is a random sample from a multivariate standard Gaussian distribution (i.e., with identity covariance matrix and zero mean vector).
For MCMC methods, as seen above, the acceptance criterion is based on log-ratio $\alpha$.
For MH-MCMC (described in Algorithm \ref{alg:MHMCMC}), it is:
\begin{equation}\label{eqn:acceptance MCMC}
    \alpha = \mathcal{J}(\bfpsi_{k})-\mathcal{J}(\bfpsi_{k+1})+\Delta q 
\end{equation}
The extended form of Eq. \eqref{eqn:acceptance MCMC}, using Eq. \eqref{eqn:proposal SN-MAP}, becomes:  
\begin{align} \nonumber
    \alpha=& \mathcal{J}({\bfpsi}_{k}) - \mathcal{J}({\bfpsi}_{k+1})- \frac{1}{2}||{\bfH}_\text{MAP}^{1/2}({\bfpsi}_{k} - {\bfpsi}_{k+1}+{\bfH}_\text{MAP}^{-1}\bfg_{k+1})||^2 
    \\ \nonumber
    &+ \frac{1}{2}||{\bfH}_\text{MAP}^{1/2}({\bfpsi}_{k+1} - {\bfpsi}_k+{\bfH}_\text{MAP}^{-1}\bfg_k)||^2 
    \\ \nonumber 
    =& \mathcal{J}({\bfpsi}_{k}) -\mathcal{J}({\bfpsi}_{k+1}) + ({\bfpsi}_{k+1} - {\bfpsi}_k)^\top(\bfg_k+\bfg_{k+1})+\frac{1}{2} ||\bfH^{-1/2}_\text{MAP} \bfg_k||^2 -\frac{1}{2}||\bfH^{-1/2}_\text{MAP} \bfg_{k+1}||^2 
\end{align}

\subsubsection{MALA}
Following Eq. \eqref{eqn:proposal MALA MAP} the updating step of MALA can be derived as:
\begin{equation}\label{eqn:}
    {\bfpsi}_{k+1} = {\bfpsi}_k - \tau {\bfH}_\text{MAP}^{-1} \bfg_k + \sqrt{(2\tau)} {\bfH}_\text{MAP}^{-1/2} \bfu  
\end{equation}
where $\bfu$ is defined as above. 
Considering $\sqrt{(2\tau)} = \Delta t$, the above equation can be re-written as:
\begin{equation}\label{eqn: updating mala}
    {\bfpsi}_{k+1} = {\bfpsi}_k -\frac{\Delta t^2}{2} {\bfH}_\text{MAP}^{-1} \bfg_k + \Delta t {\bfH}_\text{MAP}^{-1/2} \bfu
\end{equation}
The extended form of the log ratio for the acceptance criterion for MALA method, considering Eq. (\ref{eqn:proposal MALA MAP}), can be derived as:
\begin{align}\label{eqn:MALA acceptance}
    \alpha=& \mathcal{J}({\bfpsi}_{k}) - \mathcal{J}({\bfpsi}_{k+1})- \frac{1}{4\tau}||{\bfH}_\text{MAP}^{1/2}({\bfpsi}_{k} - {\bfpsi}_{k+1}+\tau{\bfH}_\text{MAP}^{-1}\bfg_{k+1})||^2 
    \\ \nonumber
    &+ \frac{1}{4\tau}||{\bfH}_\text{MAP}^{1/2}({\bfpsi}_{k+1} - {\bfpsi}_k+\tau{\bfH}_\text{MAP}^{-1}\bfg_k)||^2 
    \\ \nonumber 
    =& \mathcal{J}({\bfpsi}_{k}) -\mathcal{J}({\bfpsi}_{k+1}) + \frac{1}{2}({\bfpsi}_{k+1} - {\bfpsi}_k)^\top(\bfg_k+\bfg_{k+1})+\frac{\tau}{4} ||\bfH^{-1/2}_\text{MAP} \bfg_k||^2 -\frac{\tau}{4}||\bfH^{-1/2}_\text{MAP} \bfg_{k+1}||^2
\end{align}

\subsubsection{H-HMC}
Similarly, the updating step of H-HMC method using the Hessian information at the MAP point and leapfrog algorithm \ref{alg:HMC} can be written as:
\begin{eqnarray}\label{eqn:leapfrog}
    &\bfp_{k+\frac{1}{2}} = \bfp_k -\frac{\Delta t}{2} \bfg_k \\ \nonumber
    &{\bfpsi}_{k+1} = {\bfpsi}_k + \Delta t {\bfH}_\text{MAP}^{-1} \bfp_{k+\frac{1}{2}} \\ \nonumber
    &\bfp_{k+1} = \bfp_{k+\frac{1}{2}} -\frac{\Delta t}{2} \bfg_{k+1}
\end{eqnarray}
where $\bfp_k$ is a random sample from a Gaussian distribution with zero mean and covariance ${\bfH}_\text{MAP}$. 
Therefore, the updating step is as Eq. \eqref{eqn: updating mala}.
The extended form of the log ratio, for H-HMC method, considering Eq. \eqref{eqn:Hamiltonian}, is:
\begin{align}\label{eqn:hhmc acceptance}
    \alpha = \mathcal{J}({\bfpsi}_k) + \frac{1}{2} ||\bfH_\text{MAP}^{-1/2} \bfp_k||^2 - \mathcal{J}({\bfpsi}_{k+1}) - \frac{1}{2} ||\bfH_\text{MAP}^{-1/2} \bfp_{k+1}||^2
\end{align}
where from the leapfrog algorithm (Eq. \eqref{eqn:leapfrog}) we can define:
\begin{align}\label{eqn:hhmc leapfrog}
    \bfp_{k} = \frac{(\bfpsi_{k+1}-\bfpsi_k)^\top}{\Delta t} \bfH_\text{MAP} +\frac{\Delta t}{2}\bfg_k\\\nonumber
    \bfp_{k+1} = \frac{(\bfpsi_{k+1}-\bfpsi_k)^\top}{\Delta t} \bfH_\text{MAP} -\frac{\Delta t}{2}\bfg_{k+1}
\end{align}
and by substituting Eq. \eqref{eqn:hhmc leapfrog} in Eq. \eqref{eqn:hhmc acceptance} we derive:
\begin{align} \nonumber
    \alpha = \mathcal{J}({\bfpsi}_k) - \mathcal{J}({\bfpsi}_{k+1}) + \frac{1}{2}({\bfpsi}_{k+1} - {\bfpsi}_k)^\top(\bfg_k+\bfg_{k+1})+\frac{\Delta t^2}{8}||\bfH^{-1/2}_\text{MAP} \bfg_{k}||^2-\frac{\Delta t^2}{8}||\bfH^{-1/2}_\text{MAP} \bfg_{k+1}||^2
\end{align}
which is similar to the acceptance criterion of MALA method (Eq. \eqref{eqn:MALA acceptance}).

We summarize the important points from the comparison of Hessian-based sampling methods.
The above analysis proves that, with a proper choice of positive-definite matrix $\bfB$ in the MALA method and Mass matrix $\bfM$ in the HMC method, the updating step and acceptance coefficient of these two methods are the same.
Choice of Mass matrix $\bfM$ in HMC method and of positive-definite matrix $\bfB$ in MALA method plays a crucial role in tuning and performance of HMC and MALA methods, respectively.
In the above analysis, we propose using the Hessian information at the MAP point to tune HMC and MALA methods, which can significantly improve the performance of these methods compared to standard sampling methods. 
Comparison of these three Hessian-based methods shows that, although the SN-MAP method performs perfectly for high-dimensional Gaussian distributions, H-HMC and MALA methods with a variable step size can perform better for exploring non-Gaussian parameter spaces.
These methods will be compared numerically in Section \ref{sec:Numerical Results}.

\section{Poroelastic Forward Model}\label{sec:Forward model}

In this section, we present the forward model, based on a continuum energetic formulation of poroelasticity \cite{karimi2022energetic}, to predict the water pore pressure $p$ in the soil as a function of the permeability $\kappa$ and accounting for fluid transport in a deformable medium.
Relating this to the notation of Section \ref{sec:Bayesian framework}, the observation $i$ is a noisy pressure measure $y_i=p^i_{obs}$, and the uncertain field is $\theta(x)=-\log_e \kappa(x)$, where $\kappa$ has the units \SI{}{m^2}.
Below, we describe the governing PDEs of this model.

The governing equations of the porous medium, wherein the solid matrix is incompressible, can be written as:
\begin{align}
    \frac{\dm \trace(\bfvarepsilon)}{\dm t} + \frac{\dm {p}}{\dm t}\frac{\phi_f}{K_f} = -\frac{1}{\rho_f}\divergence\bfQ \\ \label{eqn: 4.2}
    \divergence \bfsigma + \bfb=\bf0
\end{align}
where $\epsilon$, $\phi_f$, $K_f$, and $\gamma$ are the strain tensor, fluid volume fraction, fluid bulk modulus, and dynamic viscosity of fluid, respectively. 

The flux vector $\bfQ$ is defined as:
\begin{equation}
    \bfQ = -\frac{\kappa}{\gamma}\rho_f\left( \nabla p + P_f\bfg \right) 
\end{equation}
where $\rho_f$ and $P_f$ are the true fluid density and fluid density, respectively, with the relation $P_f = \phi_f\rho_f$. In Eq. (\ref{eqn: 4.2}) $\bfsigma$ is the Cauchy stress tensor, which is defined as:
\begin{equation}
    \bfsigma = (1-\phi_{f})\big( 2\mu\bfvarepsilon + \lambda \trace(\bfvarepsilon)\bfI \big) - \phi_f p \bfI 
\end{equation}
where $\mu$ and $\lambda$ are Lame constants, and the body force is defined as $\bfb := (P_{s}+P_{f})\bfg$. where $P_s$ is the density of solid phase.

\section{Numerical Results and Discussion}\label{sec:Numerical Results}

In this section, we illustrate the results of applying the Bayesian formulation presented in Sections \ref{sec:Bayesian framework}-\ref{sec:Sampling methods}, and we compare the performance of Hessian-based sampling methods, to different examples related to the processing of pore pressure data to infer permeability.
We compare the sampling methods on a Gaussian target distribution in Section \ref{sec:Gaussian Posterior Distribution}, on a non-Gaussian target distribution in Section \ref{sec:Nearly-Gaussian Posterior Distribution},
on the posterior distribution defined by a likelihood function embedding the coupled poroelastic model of Section \ref{sec:Forward model} in Section \ref{sec:Inferring Permeability from Pressure Data}. This latter analysis is implemented in \textit{FEniCS}, and the quasi-Newton solver from the \textit{dolfin-adjoint} package is used to solve the nonlinear minimization problem. 

\subsection{Gaussian Posterior Distribution}\label{sec:Gaussian Posterior Distribution}
In this example, we consider an analytical Gaussian posterior distribution with the covariance matrix $\bfSigma$ and mean $\bfm$. We applied four sampling methods (i) MH-MCMC, (ii) HMC, (iii) SN-MCMC, and (iv) H-HMC.
The acceptance rate for SN-MCMC method is one as, considering Eqs. \eqref{eqn:modified density MCMC} and \eqref{eqn:acceptance MCMC}, it can be shown that, when the distribution is Gaussian, $q(\bfpsi_k,\bfz)$ is identical to $\omega (\bfz)$. Consequently, the log-ratio is:
\begin{equation}
    \alpha = \omega(\bfz)-\omega(\bfpsi_k)+q(\bfz, \bfpsi_k)-q(\bfpsi_k, \bfz) = 0
\end{equation}
and the acceptance coefficient becomes:
\begin{equation}\label{eqn:acc_SN-MAP}
    a = \min \Big\{ 1, \exp{(\alpha)} \Big\} = 1
\end{equation}

\begin{figure}[ht!]
	\begin{center}
		\includegraphics[width=1.0\textwidth]{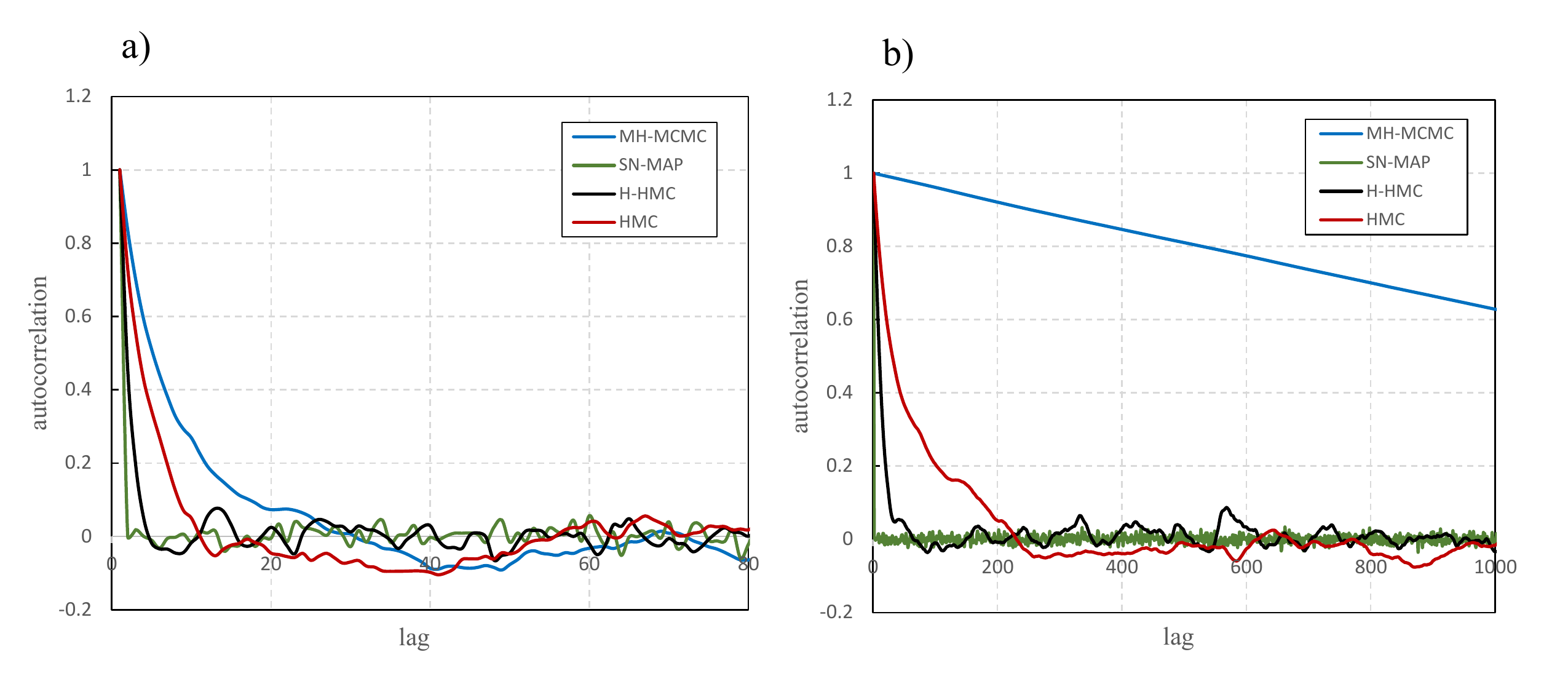}
		\caption{Autocorrelation vs. lag for different sampling methods for the a) one-dimensional case, b) high-dimensional case; autocorrelation is calculated for average value on the domain. }
		\label{fig:3}
	\end{center}
\end{figure}

Fig. \ref{fig:3} compares the autocorrelation versus lag of these four sampling methods, for a one-dimensional and for a high-dimensional Gaussian distribution. 
In the one-dimensional distribution, we assume $m = 0.5$ and $\sigma^2 = 2.0$. We generate $2,000$ samples, all the sampling processes start at $\theta = 0.5$, and the step size for MH-MCMC, HMC, and H-HMC methods is $\Delta t =1$. 

For the high-dimensional problem, we assumed a domain $\Omega$ descretized into $936$ meshpoints.
The mean value and the covariance matrix derived from the problem solved in Section \ref{sec:Inferring Permeability from Pressure Data} (the covariance matrix is as the inverse of Hessian at the MAP point).
We generate $10,000$ samples, and all the sampling processes start at the MAP (i.e. $\bfpsi=\bfm$).
The step size for MH-MCMC method is $\Delta t = 0.01$, for HMC methods is $\Delta t =0.1$, and for H-HMC method it is assumed $\Delta t = 0.3$, and considering these step sizes the acceptance rates for MH-MCMC, HMC, and H-HMC methods are approximately $85\%$, $94\%$, and $92\%$, respectively. Moreover, the acceptance rate for SN-MAP method is unitary (as shown in Eq. \eqref{eqn:acc_SN-MAP}). 

We note that, in the standard HMC and MH-MCMC sampling methods, we need to take a small step size to keep the acceptance rate high and prevent sampling from trapping in the low probable regions, which is inefficient and computationally expensive. However, using accelerated sampling methods such as H-HMC, we can choose a bigger step size while keeping the acceptance rate high, which can significantly decrease the number of samples we need to explore the parameter space.

\begin{figure}[ht!]
	\begin{center}
		\includegraphics[width=1.0\textwidth]{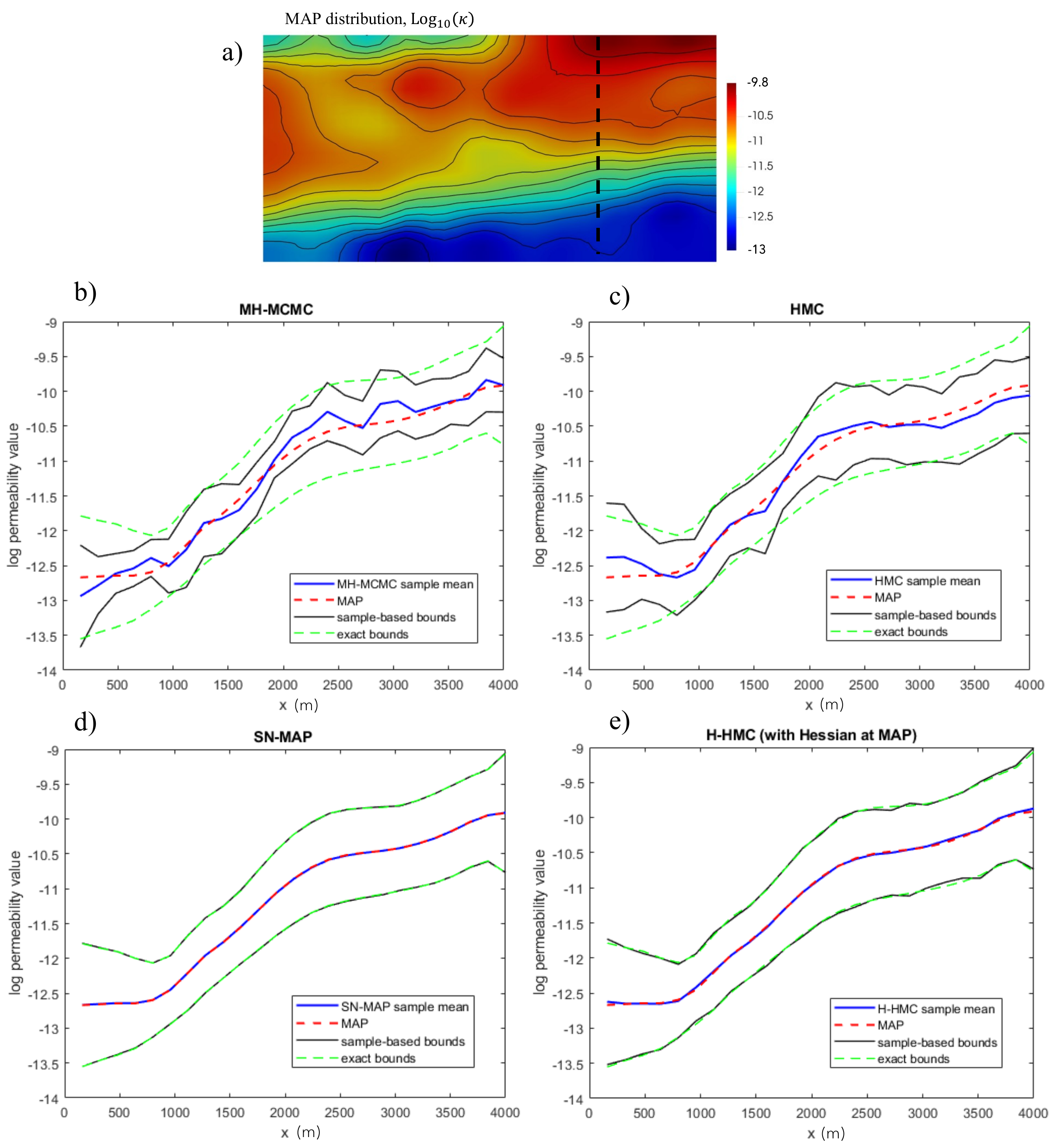}
		\caption{a) MAP point, b) interval for MH-MCMC, c) interval for HMC, d) interval for SN-MAP, e) interval for H-HMC method, for the Gaussian posterior distribution.}
		\label{fig:5}
	\end{center}
\end{figure}

As it can be seen in Fig. \ref{fig:3}, for Hessian-based sampling methods (SN-MCMC and H-HMC) the autocorrelation in both one-dimensional and high-dimensional settings quickly goes to zero, which demonstrates a faster convergence in comparison to MH-MCMC and HMC methods.
Also, comparison of H-HMC and HMC methods in high-dimension shows that we can use a bigger step size for the H-HMC method without decreasing the acceptance rate, and this increases the convergence speed.
Moreover, comparison of the autocorrelation functions for the one-dimensional and high-dimensional cases shows that the performance of MH-MCMC method decreases significantly with the dimension of the problem.  
For the high dimensional setting, Fig. \ref{fig:5} shows the $95\%$ credible interval and the sample average along the section indicated by the dashed line in the domain, based on $10,000$ samples.
The results show, the credible interval and sample average for SN-MAP and H-HMC methods matches well with the analytical solution, which are indistinguishable from the analytical solutions.
However, standard sampling methods as MH-MCMC and HMC need more samples to reach the same level of accuracy (Fig. \ref{fig:5}).

Table \ref{tab:4} represents a summary of the analysis, via correlation time $\tau$ defined as $\tau = 1 + 2\sum^\infty_{t=1} \rho_t$.
where $\rho_t$ is the correlation between two chains with lag equal to $t$.
$N_\text{eff}$ is the effective sample size which is $(N/\tau)$, and SE is the standard error which is defined as $\Tilde{\sigma}/N_\text{eff}$ where $\Tilde{\sigma}$ is the sample chain standard deviation.
\begin{table*}[h]
    \begin{subtable}[h]{0.53\textwidth}
    \centering
    \begin{tabular}{|c|c|c|c|}
        \hline
        Method  & $\tau$  & $N_\text{eff}$  & SE \\
        \hline
        HMC  & 1149.3  & 9  & 0.0036 \\
        H-HMC  & 62.5  & 160  & 0.00027 \\
        SN-MAP  & 1  &10,000  & $1.4\times10^{-5}$ \\
        \hline
    \end{tabular}
    \caption{Summery of convergence analysis of different sampling methods for high-dimensional normal distribution.}
    \label{tab:4}
    \end{subtable}
    \begin{subtable}[h]{0.42\textwidth}
    \centering
    \begin{tabular}{|c|c|c|} 
        \hline
        Case \# & $\sigma_{\epsilon}$ & meas. \\
        \hline
        1  & $0.5$ MPa  & 60\\
        2  & $2$ MPa  & 60\\
        3  & $4$ MPa & 60\\
        4  & $4$ MPa  & 20\\
        5  & $10$ MPa  & 20\\
        \hline
    \end{tabular}
    \caption{Cases with different uncertainty levels, level of noise ($\sigma_\epsilon$) and number of measurements.}
    \label{tab:2}
    \end{subtable}
    \caption{Summery analysis of high-dimensional normal example.}
\end{table*}
The table shows how the SN-MAP sampling method has the lowest standard error and it is the most efficient method for the high-dimensional normal distribution.
We do not consider the MH-MCMC method in this analysis, since the sampling chains of this method are yet far from convergence after 
$10,000$ iterations. 

\begin{figure}[h]
	\begin{center}
		\includegraphics[width=0.6\textwidth]{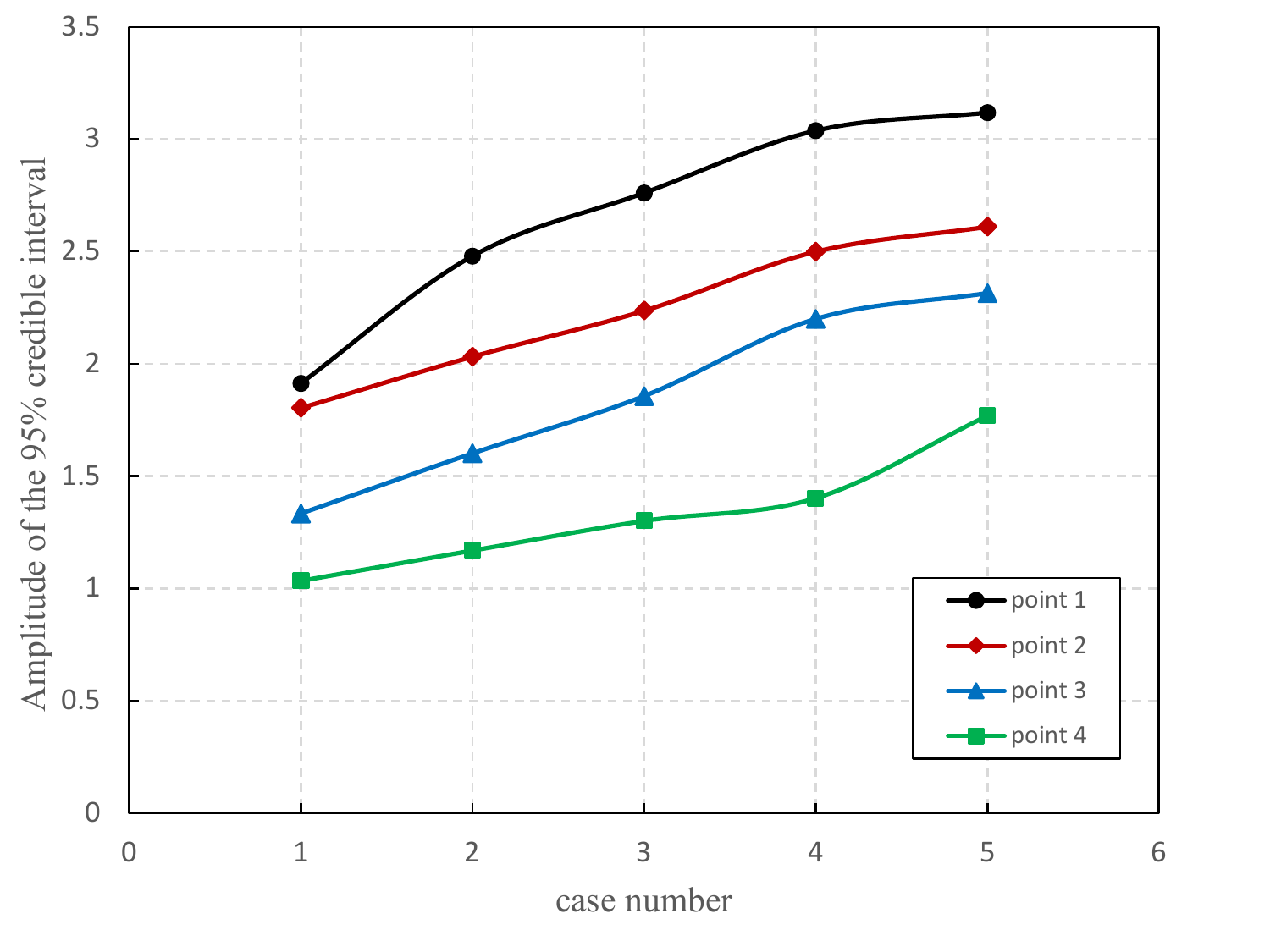}
		\caption{Amplitude of the $95\%$ credible interval vs. measurement uncertainty for the Gaussian posterior distribution, using the H-HMC method, at point 1:($1371.4, 0$), point 2:($0,160$), point 3:($914.3, 320$), and point 4:($4114.3,1280$).}
		\label{fig:12-3}
	\end{center}
\end{figure}

Table \ref{tab:2} represents $5$ different cases, with different levels of uncertainty, to investigate the effect of measurement number and noise level on the credible interval amplitude.
We increase the noise level by decreasing the number of measurements and increasing the noise level.
For each case, the MAP point and Hessian at the MAP point is computed.
Then, considering the normal distribution approximating the posterior distribution at the MAP, we take $10,000$ samples using the H-HMC method and compute the $95\%$ credible interval at four random points in the domain, for each case. The results are reported in Fig. \ref{fig:12-3}.
As we expected, the results show that by increasing the uncertainty level, the amplitude of the credible interval increases (Fig. \ref{fig:12-3}).
\subsection{Nearly-Gaussian Posterior Distribution}\label{sec:Nearly-Gaussian Posterior Distribution}
In this section, we apply the sampling methods to a log-normal target distribution. 
The objective function is defined as:
\begin{align}
    \mathcal{J}(\bfpsi) = \frac{1}{2} ||\bfLambda^{1/2}\big( \log (\bfpsi) -\bfm_l \big)||^2  - \log (\Pi^N_{i=1} \psi_i^{-1})
\end{align}
 where $\bfLambda = \bfSigma^{-1}_l$. $\bfSigma_l$ and $\bfm_l$ are covariance matrix and mean vector inputs for the log-normal distributions, respectively. The local gradient and Hessian of the objective function can be derived as follows:
\begin{align}\label{eqn:gradient log-normal}
    \parderiv{\mathcal{J}(\bfpsi)}{\psi_j} = \psi_j^{-1} \Big[ \bfn_j^T \bfLambda \big(  \log (\bfpsi) -\bfm_l \big) + 1 \Big]
\end{align}

where $\bfn_j=\{n_j^i\}_{i=1}^N$ is a vector with components $n_j^i = 0,~ i \neq j$ and $n_j^i = 1, ~ i=j$.

\begin{equation}
    \frac{\partial^2 \mathcal{J}(\bfpsi)}{\partial \psi_j \partial \psi_i} =\left\{\begin{array}{ll}
    -\psi_j^{-2} \big[ \bfn_j^T \bfLambda \big(  \log (\bfpsi) -\bfm_l \big) + 1 + \Lambda_{jj}\big] & \quad\text{if } i = j\\
    \psi_j^{-1} \big[ \Lambda_{ji} \psi_i^{-1} \big] & \quad\text{if } i \neq j 
    \end{array}\right.
\end{equation}
Based on Eq. \eqref{eqn:gradient log-normal} the MAP point is $\bfpsi_\text{MAP} = \exp(\bfm_l - \bfSigma_l\bf1)$.
We discuss the performance of standard and accelerated sampling methods for one-dimensional and high-dimensional log-normal distributions. 
In the one-dimensional case, we assume $m_l = 0.5$ and $\sigma_l^2 = 2.0$.
We generate $4,000$ samples, and the step size for MH-MCMC, HMC, and H-HMC methods is $\Delta t =1$.
All the sampling processes start from the mode of log-normal distribution, $\exp (m_l -\sigma_l^2)$.

As a second setting, we consider a high-dimensional log-normal distribution. The domain and distribution are assumed as Section \ref{sec:Gaussian Posterior Distribution}, and the MAP point $\bfpsi_\text{MAP} = \exp(\bfm_l - \bfSigma_l\bf1)$ is assumed similar to the MAP point of previous example (Fig. \ref{fig:5}). Similarly, the step size for MH-MCMC method is $\Delta t = 0.01$, for HMC method is $\Delta t =0.1$, and for H-HMC method it is $\Delta t = 0.3$. The correspondence acceptance rates after taking $10,000$ samples for MH-MCMC, HMC, and H-HMC methods are $82\%$, $94\%$ and $92\%$, respectively.
In addition, the acceptance rate for SN-MAP method is approximately $8.3\%$, which is significantly lower than that of the SN-MAP sampling method for high-dimensional Gaussian distribution.

\begin{figure}[ht!]
	\begin{center}
		\includegraphics[width=1.0\textwidth]{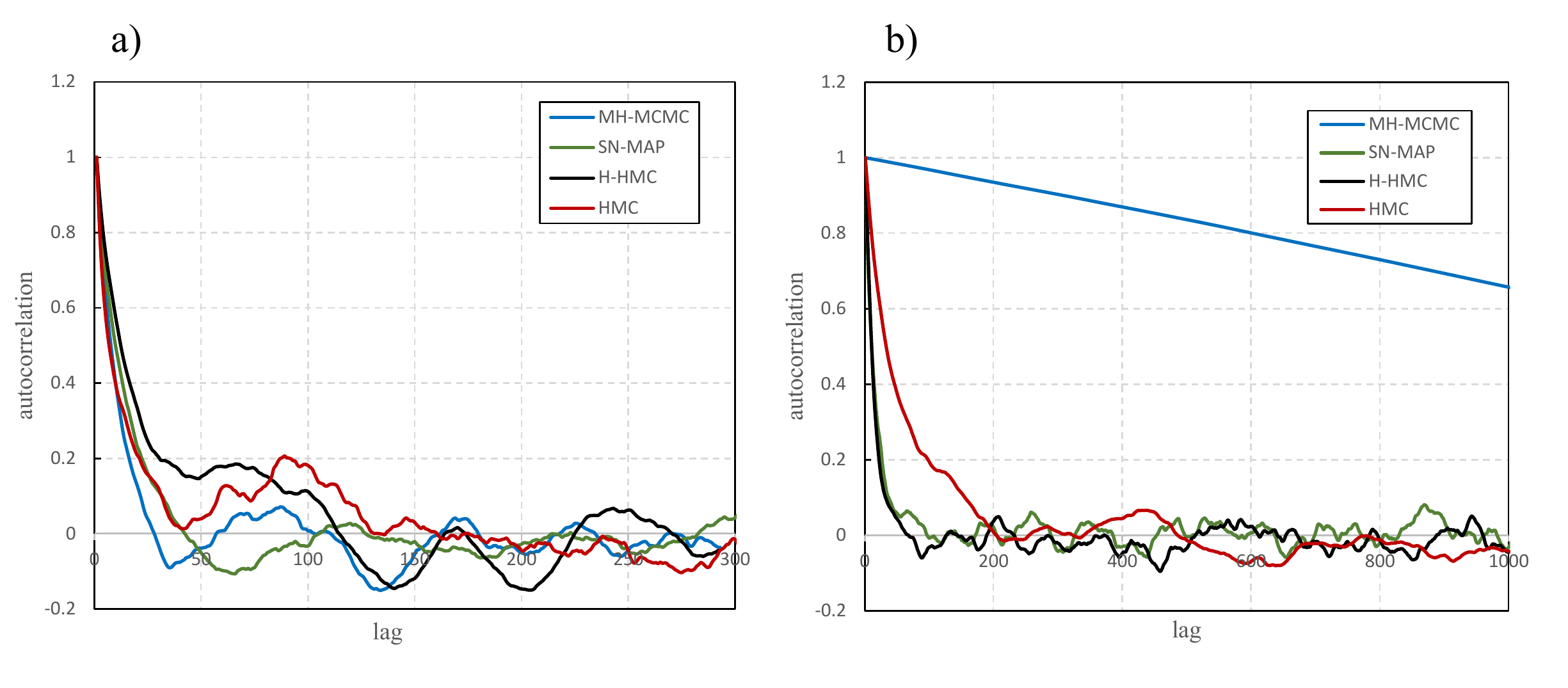}
		\caption{Autocorrelation vs. lag of different sampling methods for the a) one-dimensional case, b) high-dimensional case; autocorrelation is calculated for average value on the spatial domain.}
		\label{fig:6}
	\end{center}
\end{figure}

Fig. \ref{fig:6} represents the autocorrelation function versus lag, for one-dimensional and high-dimensional settings. The results show that the performance of the four methods in one-dimensional distribution is quite similar.
In addition, as it is shown in Fig. \ref{fig:6} the convergence speed of SN-MAP method and H-HMC method with the Hessian information at the MAP point in high-dimensional distribution are significantly faster than standard MH-MCMC method.
We also, considered another example, where we take sample from a log-normal distribution and the MAP point is assumed $\bfpsi_\text{MAP}=\exp(\bfm)$. where $\bfm$ is assumed similar to the MAP point of the normal example. We have reported the results in Appendix \ref{sec:Appendix D}.

\begin{figure}[ht!]
	\begin{center}
		\includegraphics[width=1.0\textwidth]{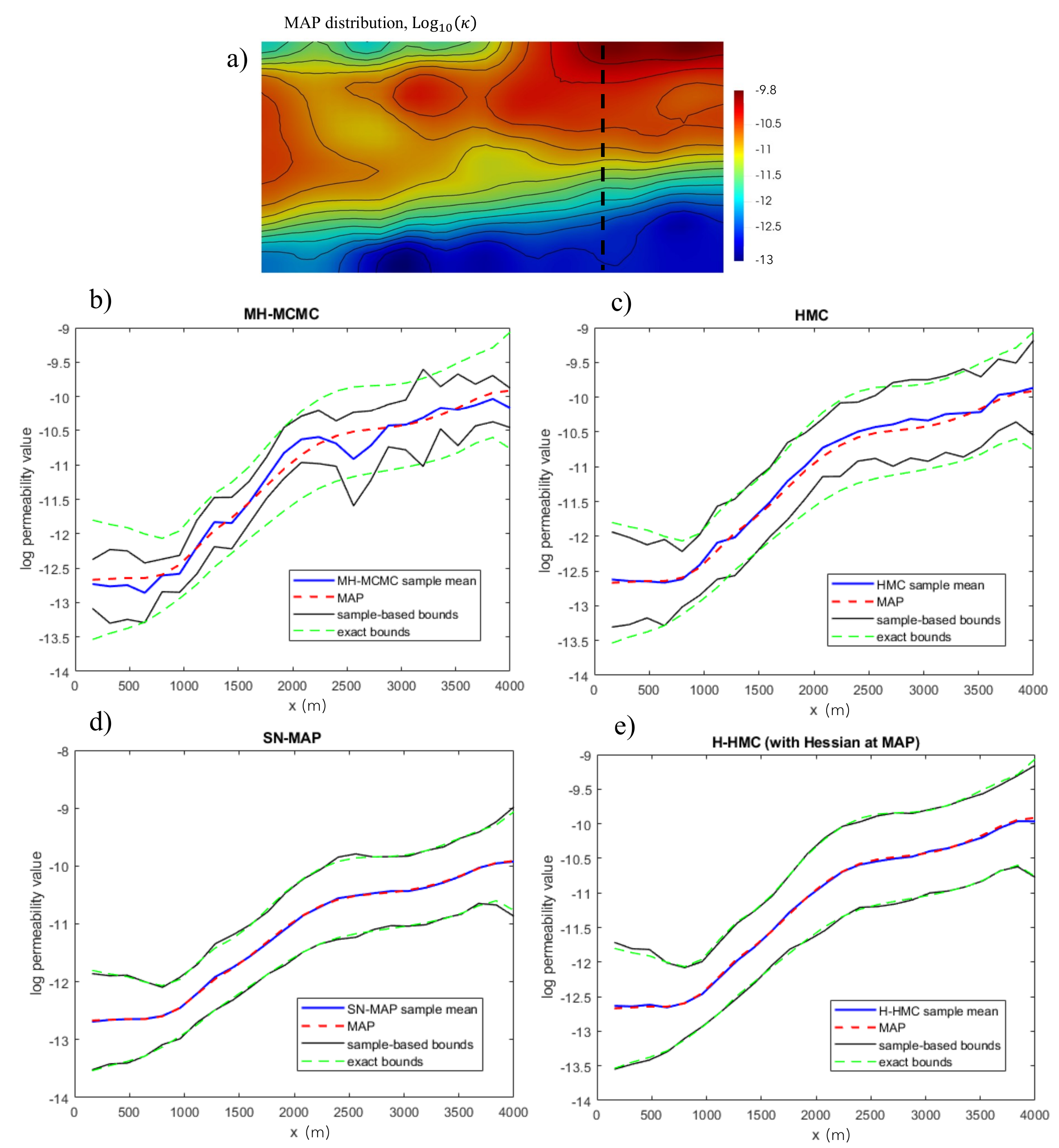}
		\caption{a) MAP point, b) interval for MH-MCMC, c) interval for HMC, d) interval for SN-MAP, e) interval for H-HMC method, for the nearly-Gaussian posterior distribution.}
		\label{fig:10}
	\end{center}
\end{figure}

Fig. \ref{fig:10} presents the $95\%$ credibility interval and sample average along the dashed line for MH-MCMC, HMC, SN-MAP, and H-HMC methods. As the results show, the HMC method converges and explores the parameter space faster than the MH-MCMC method. H-HMC and SN-MAP methods convergence speeds are significantly higher than standard sampling methods. After $10,000$ samples, the sample credible interval matches well with the exact credible interval for both H-HMC and SN-MAP methods. 

Table \ref{tab:5} shows the summary of convergence analysis for the high-dimensional log-normal distribution, after taking $10,000$ samples using MH-MCMC, HMC, and H-HMC sampling methods. Results of table \ref{tab:5} show a decrease in the value of $N_\text{eff}$ compares to applying similar sampling methods to high-dimensional normal distribution (table \ref{tab:4}), which proves that taking samples from the log-normal distribution is more difficult.

Moreover, comparison of SE values in tables \ref{tab:4} and \ref{tab:5} shows that although the SN-MAP method works well for normal distributions and it has the lowest estimation error value, for log-normal distribution using the H-HMC method is more efficient, and the estimation error is lower. 
\begin{table*}[h]
    \begin{subtable}[h]{0.45\textwidth}
    \centering
    \begin{tabular}{|c|c|c|c|} 
        \hline
        Method & $\tau$ & $N_\text{eff}$ & SE \\
        \hline
        HMC & 1458.1 & 7 & 0.0062 \\
        H-HMC & 70.7 & 141 & 0.0003 \\
        SN-MAP & 87.7 & 114 & 0.0004 \\
        \hline
    \end{tabular}
    \caption{Summery of convergence analysis of different sampling methods for high-dimensional log-normal distribution.}
    \label{tab:5}
    \end{subtable}
    \begin{subtable}[h]{0.45\textwidth}
    \centering
    \begin{tabular}{|c|c|c|} 
        \hline
        Case \# & KLD val. & acceptance $\%$ \\
        \hline
        1  & 2.47  & 8.9 \\
        2  & 3.15  & 8.3 \\
        3  & 5.04  & 7.3 \\
        4  & 13.55  & 4.3 \\
        5  & 69.03  & 1.7 \\
        6  & 195.02  & 0.3 \\
        \hline
    \end{tabular}
    \caption{Results of SN-MAP sampling method for different log-normal distributions.}
    \label{tab:3}
    \end{subtable}
    \caption{Summery analysis of high-dimensional log-normal example.}
\end{table*}

Table \ref{tab:3} presents the Kullback-Leibler divergence (KLD) values between normal and log-normal distributions and the acceptance rates for different log-normal distributions. This analysis considers different log-normal distributions and calculates the KLD value, which measures the difference between each log-normal distribution and its corresponding normal approximation at the MAP.
For each log-normal distribution, we run the SN-MAP sampling method and take $10,000$ samples. As the results show, by increasing the KLD value, the acceptance rate decreases, which shows that, for distributions that are not close to Gaussian, generating samples is more expensive as the acceptance rate is lower.
The details related to computing the KLD between normal and log-normal distributions is presented in Appendix \ref{sec:Appendix C}.

\subsection{Non-Gaussian Posterior Distribution: Inferring Permeability from Pressure Data}\label{sec:Inferring Permeability from Pressure Data}
\begin{figure}[h]
	\begin{center}
		\includegraphics[width=1.0\textwidth]{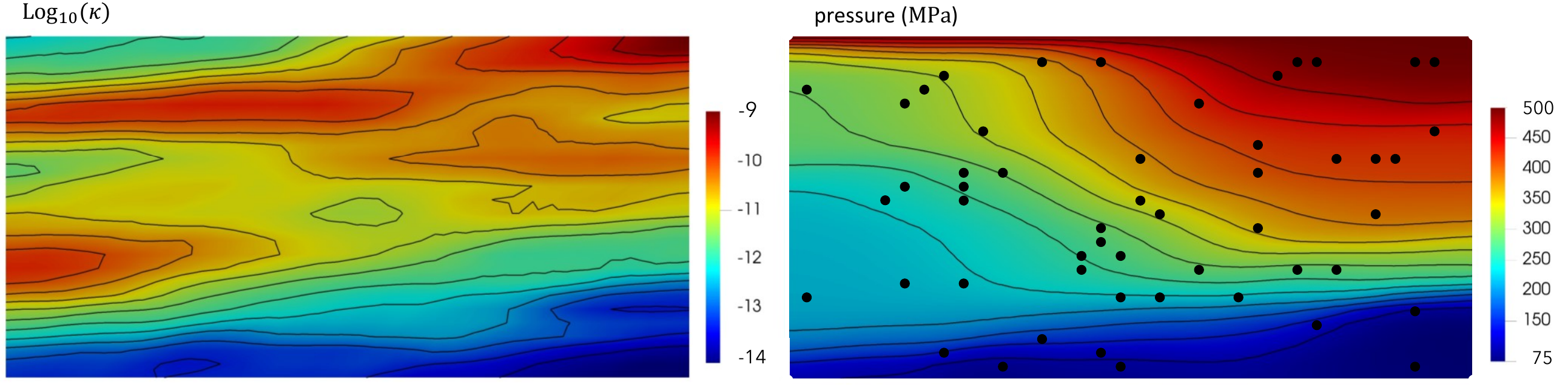}
		\caption{(left) target permeability distribution $\log_{10} (\kappa)$, (right) synthetic pointwise pressure $p$ observations 10 days after start of the injection.}
		\label{fig:11}
	\end{center}
\end{figure}

In this section, we infer the permeability field $\kappa$ from the sparse pore pressure data.
The PDE governing equations of forward model has described in section \ref{sec:Forward model}. 
The properties of assumed underground layer are listed in table \ref{tab:1}. 
The boundary conditions are assumed as below:
\begin{gather*}
    \text{top~BC}: ~ p = \const ~~,~~
    \text{bottom ~ BC}: ~ ~\bfdelta = \bf0
\end{gather*}
The initial porosity is assumed $\phi_{0f} = 0.2$. We consider a $8,000 \times 4,000$ deep underground layer with the constant fluid pressure of $p=500$ \SI{}{M\pascal} on the top boundary.
\begin{table*}[htbp]
    \centering
    \begin{tabular}{|l|l|}
        \hline
        Property & Value\\
        \hline
        Solid phase Lame constant, ${\lambda}$ \qquad \qquad & \SI{40}{\mega\pascal}\\
        Solid phase Lame constant, ${\mu}$ & \SI{40}{\mega\pascal}\\
        Fluid bulk modulus, $K_f$ &  \SI{2270}{\mega\pascal}\\
        Water density, $\rho_f$ & \SI{1000}{\kg \per m^3}\\
        Water viscosity, $\gamma$ & \SI{0.001}{\pascal\cdot s}\\
        Solid density, $\rho_s$ & \SI{2000}{\kg \per m^3}\\
        \hline
    \end{tabular}
    \caption{Properties of solid and fluid phases in the poroelastic model, see Section \ref{sec:Forward model}.}\label{tab:1}
\end{table*}
Fig. \ref{fig:11} shows the assumed (hidden) permeability field $\kappa$, used to generate artificial measures by running the forward model.
In this example, we generated $52$ synthetic observations of pore pressure, affected by random noise with standard deviation $\sigma_{\epsilon} = 1.0$ \SI{}{M\pascal}. The observations are assumed in a single time horizon 10 days after start of the injection.

As discussed in Section \ref{sec:Bayesian framework}, the prior covariance is defined based on the Laplacian operator $\mathcal{A}^{-1} = -\gamma\bfPsi{\Delta} + \delta\bfI$, with $\gamma = 0.5$, $\delta = 5\times 10^{-3}$ (\SI{}{1/km^2}), $a = 0.018$, $b = 0.97$, and $\beta = 1.017\pi$ (the prior parameters are defined in Section \ref{sec:choice of prior}).

\subsubsection{Computing the MAP Point}
\begin{figure}[ht!]
	\begin{center}
		\includegraphics[width=0.6\textwidth]{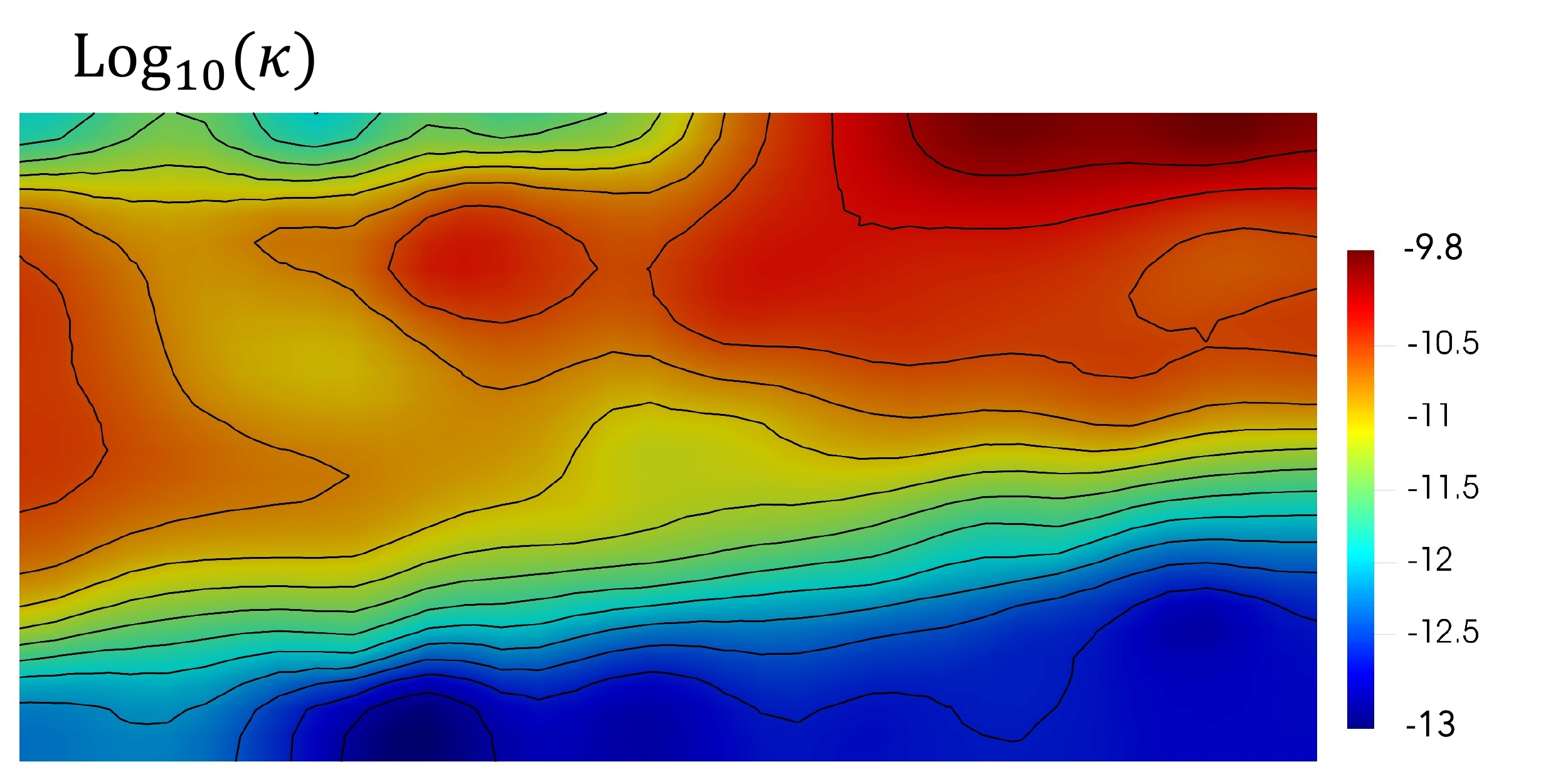}
		\caption{The MAP point field with $\sigma_{\epsilon}=1.0$ \SI{}{M\pascal}.}
		\label{fig:12}
	\end{center}
\end{figure}

\begin{figure}[ht!]
	\begin{center}
		\includegraphics[width=1.0\textwidth]{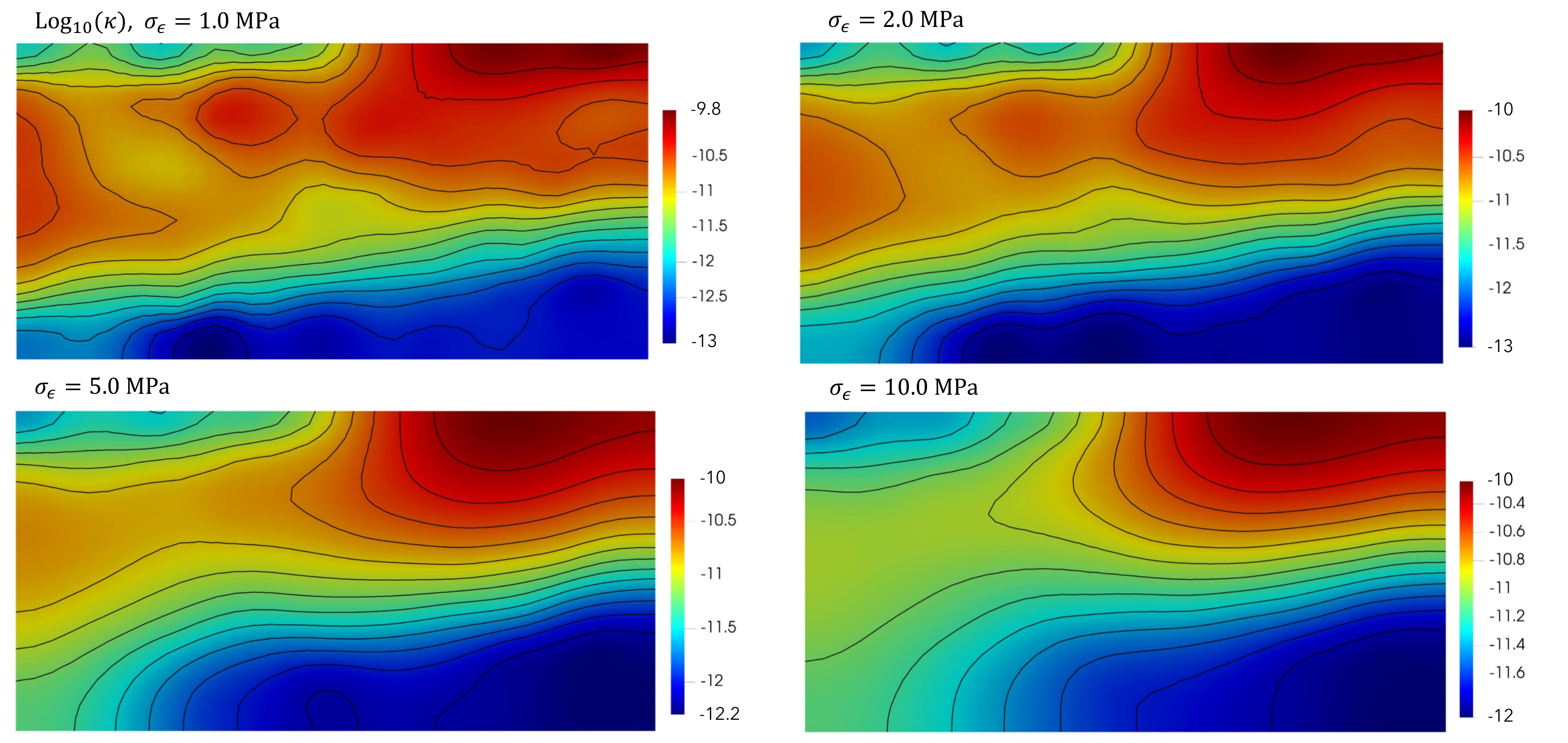}
		\caption{The MAP field considering different noise levels with $52$ observations.}
		\label{fig:12-1}
	\end{center}
\end{figure}

\begin{figure}[ht!]
	\begin{center}
		\includegraphics[width=1.0\textwidth]{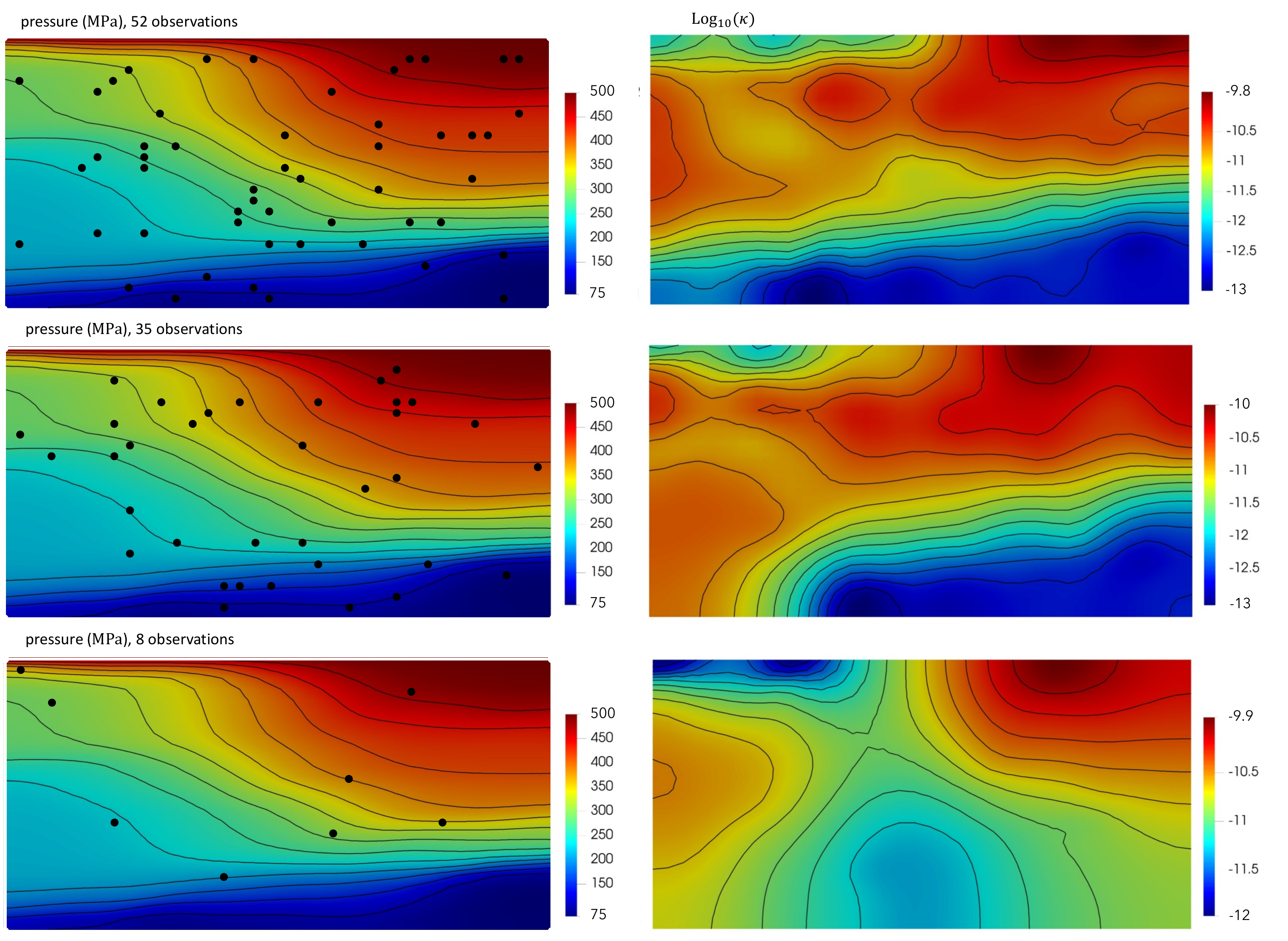}
		\caption{The MAP field considering different measurement numbers, with $\sigma_{\epsilon} = 1$ MPa.}
		\label{fig:12-2}
	\end{center}
\end{figure}

To find the MAP point, we apply a quasi Newton algorithm (BFGS) to solve the nonlinear minimization problem.
The detailed formulation of Lagrangian method is presented in Appendix \ref{sec:Appendix B}. The prior mean is assumed constant ($m_{\pi} = 33$ \SI{}{m^2}).
We start with an initial guess for the field and by applying the Newton method, we iteratively update the field value.
Fig. \ref{fig:12} shows the MAP point (which is similar to actual permeability shown in Fig. \ref{fig:11}).
Fig. \ref{fig:12-1} shows the identified MAP points by considering different noise levels.
By increasing the level of noise $\sigma_{\epsilon}$, the prior term becomes more dominant and consequently the MAP field is smoother. 
Similarly, Fig. \ref{fig:12-2} shows the effect of number of measurements on computing the MAP point. 
By decreasing the number of observations, the MAP point becomes less informed, more smooth and far from the target distribution.

\subsubsection{Exploring the Posterior Distribution}
\begin{figure}[ht!]
	\begin{center}
		\includegraphics[width=0.55\textwidth]{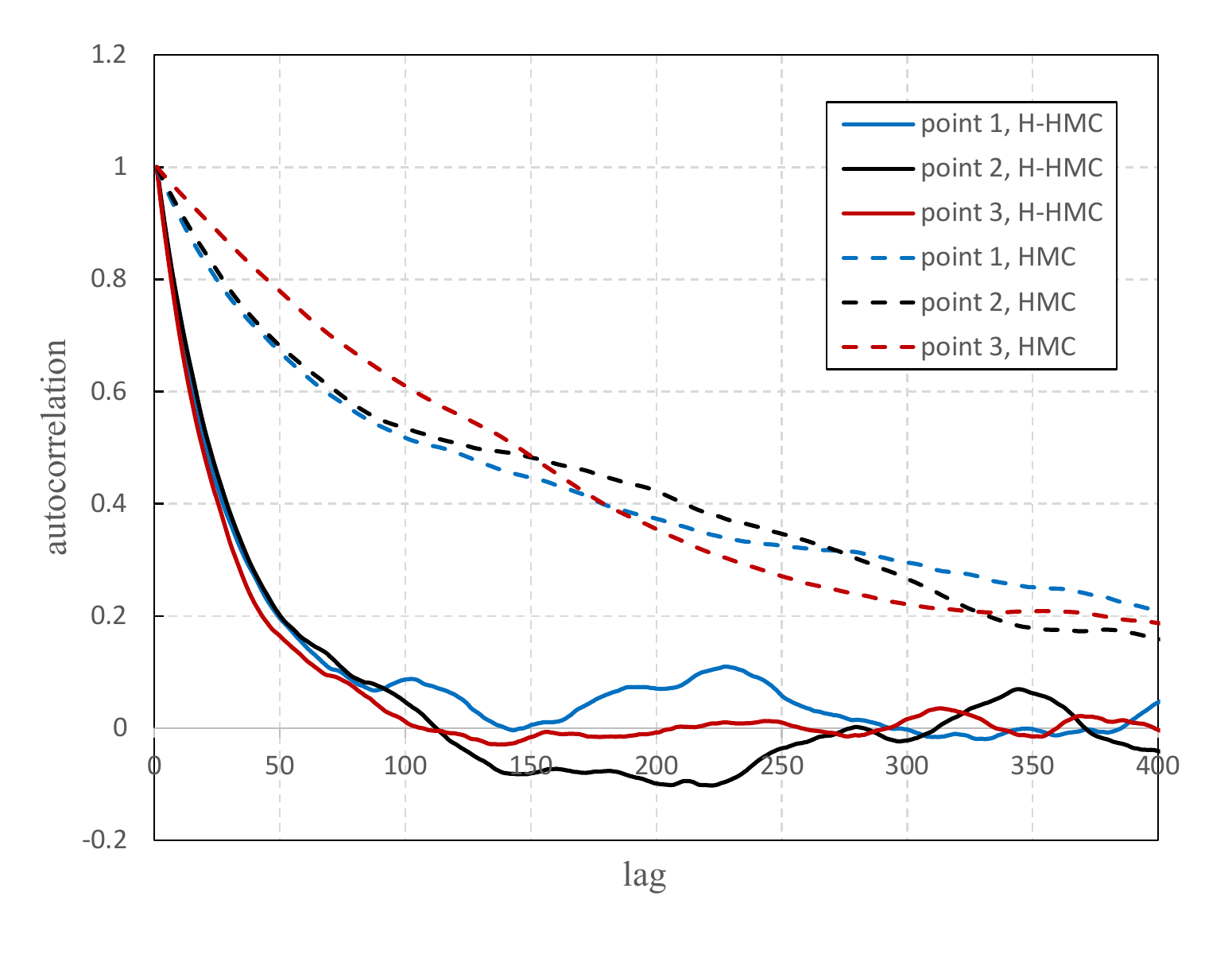}
		\caption{Autocorrelation vs. lag at 3 random points, point 1:(7314.3,800), point 2:(5028.8,2880), and point 3:(10,3200).}
		\label{fig:15}
	\end{center}
\end{figure}

We use the HMC and H-HMC methods to generate samples from the posterior distribution. In the H-HMC method, we use the Hessian information at the MAP point to improve the sampling performance. To achieve similar acceptance rates, the step size $\Delta t$ for HMC and H-HMC methods are set as $0.1$, and $0.3$, respectively. After taking $20,000$ samples for HMC, the correspondence acceptance rate and H-HMC methods are approximately $85\%$ and $83\%$, respectively. 
Fig. \ref{fig:15} shows the autocorrelation function versus lag, computed for the permeability value at three points in the domain using HMC and H-HMC methods.

\begin{figure}[ht!]
	\begin{center}
		\includegraphics[width=1.0\textwidth]{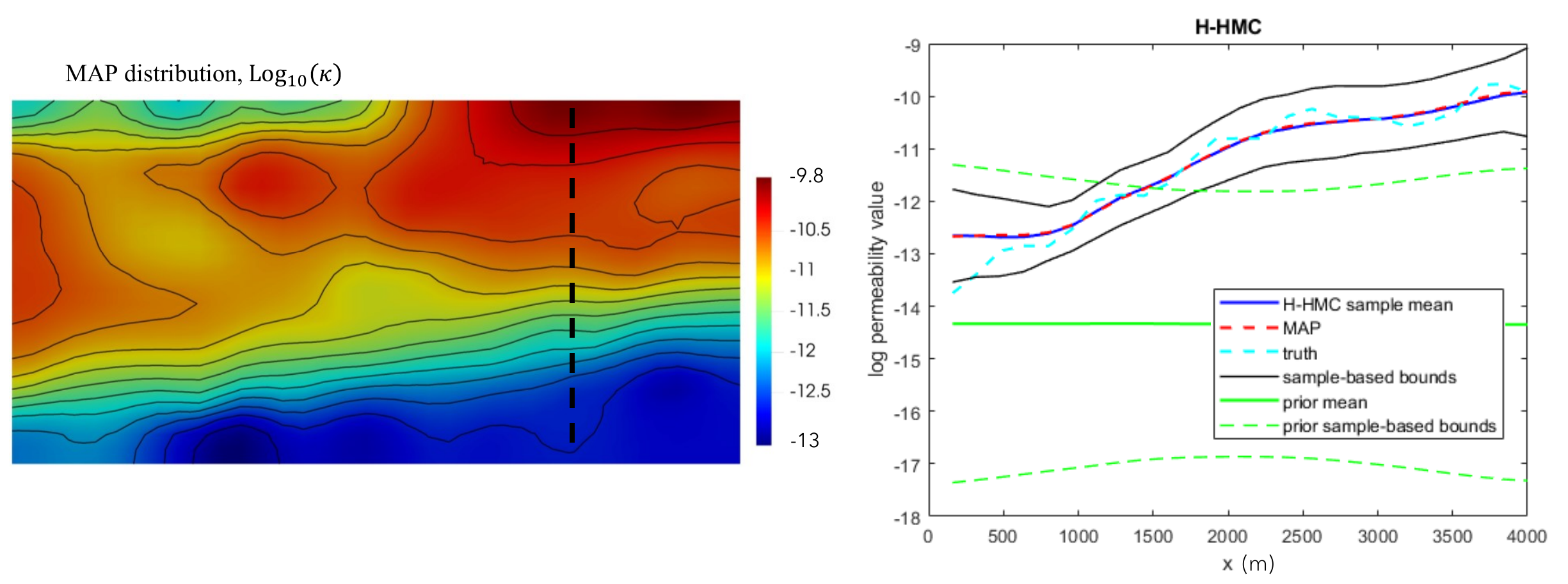}
		\caption{$95\%$ credible interval based on the H-HMC method.}
		\label{fig:16}
	\end{center}
\end{figure}

In this analysis, we don't use the MH-MCMC method due to the inefficiency and computational costs. Also, since the posterior distribution is highly nonlinear and non-Gaussian, the acceptance rate for the SN-MAP method is very low, which makes this method inefficient. 
Furthermore, Fig. \ref{fig:16} presents the credible interval of the inverse problem’s solution along the vertical dashed line, after taking $50,000$ samples using H-HMC method. The graph also shows the prior samples' mean and $95\%$ credible interval. The comparison of prior and posterior intervals shows that the posterior variance has decreased significantly, ensuring that the posterior distribution is highly influenced by observations. In addition, posterior samples' mean, MAP point and the target value are shown in Fig. \ref{fig:16}. As the results show, the posterior samples' mean and MAP point match well. 

\section{Conclusion}
We have presented a Bayesian inference framework for high-dimension inverse problems governed by PDE equations.
We used a continuous Gaussian prior distribution (based on a Laplacian-like operator) to ensure the well-posedness of the infinite-dimensional inverse problem.
The main advantage of this prior function is that it can be applied to continuous domains, and it allows simple discretization.
We implemented the inverse problem in the \textit{FEniCS} library and used an quasi-Newton solver (BFGS) of \textit{dolfin-adjoint} package to solve the minimization problem. 

We have investigated several sampling methods to describe the posterior distribution.
We discussed the complexities of sampling methods in high-dimensional parameter spaces and compared the performance of MH-MCMC and HMC sampling methods with the accelerated methods using the Hessian and gradient information, such as SN-MCMC, MALA, and H-HMC.
By considering several one-dimensional and high-dimensional problems with Gaussian and non-Gaussian posterior distributions, we showed that using the modified Hessian-based sampling methods can significantly increase the speed of convergence and exploring in high-dimensional inverse problems. However, using the local Hessian information in nonlinear high-dimensional problems is computationally expensive. We considered Hessian sampling methods using the Hessian information calculated at the MAP point to overcome this problem. The results revealed that the MAP-based Hessian sampling methods are both computationally efficient and fast in exploring high-dimensional distributions.

We also applied the developed framework to a high-dimensional inverse problem governed by poroelastic PDE equations to infer the unknown permeability distribution from the point-wise pore pressure observations.
We calculated the MAP point and the posterior credible intervals by applying the HMC and H-HMC methods, using the Hessian information at the MAP point. Our results indicate that the H-HMC method using the Hessian information at the MAP point has a better performance in exploring this non-Gaussian high-dimensional distribution.

\section*{Software Availability}

A version of the code developed for this work is available at \url{https://github.com/minakari/Bayesian}.


\section*{Acknowledgments}
We thank the National Science Foundation for support through XSEDE resources provided by Pittsburgh Supercomputing Center. 
Mina Karimi acknowledges financial support from the Scott Institute. 
Kaushik Dayal acknowledges financial support from NSF (CMMI MOMS 1635407, DMS 2108784), ARO (MURI W911NF-19-1-0245), ONR (N00014-18-1-2528), BSF (2018183), and an appointment to the National Energy Technology Laboratory sponsored by the U.S. Department of Energy. 
Matteo Pozzi acknowledges financial support from NSF (CMMI 1638327).
This work was funded (in part) by the Dowd Fellowship from the College of Engineering at Carnegie Mellon University.

\appendix

\section{Constructing an Infinite-dimensional Gaussian Prior}\label{sec:Appendix bA}

The Gaussian prior field in the infinite dimension domain can be defined via the Karhunen-Loeve expansion ~\cite{dashti2013bayesian}, as:
\begin{equation}
\label{eqn:Kahunen-Loeve expansion}
    \theta = m_\pi + \sum ^\infty_{j=1} \lambda_j \xi_j \Tilde{\phi}_j
\end{equation}
where $\{\Tilde{\phi}_j\}^\infty_{j=1}$ is a set of orthonormal basis functions of the domain, $\{\lambda_j\}^\infty_{j=1}$ is a deterministic sequence and $\{\xi_j\}^\infty_{j=1}$ is a set of independent standarly distributed normal variables $\xi_j \sim \mathcal{N} (0,1)$.
Using Eq. \eqref{eqn:Kahunen-Loeve expansion}, the covariance operator in Eq. \eqref{eqn:covarian operator} can be written as follows:
\begin{eqnarray}
    \mathcal{C}_\pi &=& \E \Big( \sum^\infty_{j=1} \sum^\infty_{k=1} \lambda_j \lambda_k \xi_j \xi_k \Tilde{\phi}_j \otimes \Tilde{\phi}_k \Big) = \Big( \sum^\infty_{j=1} \sum^\infty_{k=1} \lambda_j \lambda_k \E (\xi_j \xi_k) \Tilde{\phi}_j \otimes \Tilde{\phi}_k \Big) \nonumber\\
    &=& \sum^\infty_{j=1} \sum^\infty_{k=1} \lambda_j \lambda_k \delta_{jk} \Tilde{\phi}_j \otimes \Tilde{\phi}_k = \sum^\infty_{j=1} \lambda^2_j \Tilde{\phi}_j \otimes \Tilde{\phi}_j \nonumber
\end{eqnarray}
By orthonormality of set $\{ \Tilde{\phi}_j \}^\infty_{j=1}$, we can write:
\begin{eqnarray}
    \mathcal{C}_\pi \Tilde{\phi}_k &=& \Big( \sum^\infty_{j=1} \lambda^2_j \Tilde{\phi}_j \otimes \Tilde{\phi}_j \Big) \Tilde{\phi}_k = \sum^\infty_{j=1} \lambda^2_j \langle \Tilde{\phi}_j,\Tilde{\phi}_k \rangle \Tilde{\phi}_j 
    = \sum^\infty_{j=1} \lambda^2_j \delta_{jk} \Tilde{\phi}_k = \lambda^2_k \Tilde{\phi}_k \nonumber
\end{eqnarray}
Thus, the above calculation implies that $\big( \lambda^2_j , \Tilde{\phi}_j \big)^\infty_{j=1}$ are the eigenpairs of the covariance operator $\mathcal{C}_\pi$.  
For constructing a prior as Eq. \eqref{eqn:Kahunen-Loeve expansion} of the random field $\theta$, we need to specify the mean function $m_\pi$ and covariance operator $\mathcal{C}_\pi$.

\section{Low-rank Hessian}\label{sec:Appendix A}

The Hessian or second derivative of the objective function (Eq. \eqref{eqn:objective function discrete}) w.r.t the controlling parameter can be written as:
\begin{equation}
\label{eqn:Hessian}
    \bfH = \bfH_\text{misfit} + \bfA^{-1}
\end{equation}
where $\bfH_\text{misfit}$ is the Hessian of first term in the objective function, which shows the misfit between the observations and predicted values, and $\bfA$ is the Hessian of prior term. We can consider the Cholesky decomposition of the prior covariance as $\bfA = \bfL \bfL^\top$.
Using this decomposition, we can rewrite Eq. \eqref{eqn:Hessian} as $\bfH = \bfL^{-\top} \big( \bfL^\top \bfH_\text{misfit} \bfL +\bfI \big)\bfL^{-1}$.

where the first term $\bfL \bfH_\text{misfit} \bfL$ is more informative about the controlling parameter, and its $r-$dimensional low rank approximation can be written as $\bfL^\top \bfH_\text{misfit} \bfL \approx \bfV_r \bfD_r \bfV_r^\top$.
where $\bfV_r$ is the $n\times r$ matrix of dominant eigenvectors, and $\bfD_r$ is the $r\times r$ diagonal matrix of dominant eigenvalues. 
Therefore, the low rank approximation of $\bfH$ can be written as $\Tilde{\bfH} = \bfL^{-\top} \big( \bfV_r \bfD_r \bfV_r^\top +\bfI \big)\bfL^{-1}$.

\section{Lagrangian Formulation}\label{sec:Appendix B}

We minimize the objective function (negative log-posterior) to find the MAP point. In this example, the objective function is defined as $\mathcal{J}(\theta) = \frac{1}{2\sigma^2_{\epsilon}}\sum^n_{i=1} (p(\theta) - p^i_{obs})^2 + \frac{1}{2}||\mathcal{A}(\theta-\theta_{pr})||^2 $.

where $p(k)$, $\{p_{obs}\}^N_{i=1}$, and $\sigma_{\epsilon}^2$ are predicted values of the pore pressure, pressure measurements, and noise covariance, respectively. $\theta_{pr}$ is the prior mean. 
In addition, the Lagrangian form can be written as the summation of objective function and weak form of the governing equations:
\begin{align} \nonumber
    \mathcal{L}(\theta, p, \Tilde{p}, \bfepsilon, \Tilde{\bfu}) &= \mathcal{J}(\theta) + \int^\top_0 \int_{\Omega} (\trace(\bfepsilon)_t + p_t)\Tilde{p}\dm\bfx\dm t + \int^\top_0\int_{\Omega} \frac{e^{-\theta}}{\gamma} (\nabla p + \bfb )\cdot \nabla\Tilde{p} \dm\bfx\dm t\\ \nonumber
    &- \int^\top_0\int_{\partial\Omega} \frac{e^{-\theta}}{\gamma}(\nabla p +\bfb)\cdot \bfn \Tilde{p} \dm \bfx \dm t + \int^\top_0 \int_{\Omega} (\bfb\cdot\Tilde{\bfu} - \bfsigma \cdot \Tilde{\bfu}) \dm\bfx \dm t
\end{align}
where $\Tilde{p}$ and $\Tilde{\bfu}$ are the Lagrange multipliers for balance of mass and balance of momentum equations, respectively. $(\cdot)_t$ represents the derivative with respect to time $\frac{\dm(\cdot)}{\dm t}$. 
The optimality conditions (Karush-Kuhn-Tucker conditions) for solving the minimization problem, can be found by calculating the variation with respect to the Lagrangian function variables. We have used the \textit{dolfin-adjoint} python library to calculate the optimality conditions and find the MAP point. 

\section{Kullback-Leibler divergence between normal and log-normal distributions}\label{sec:Appendix C}

We assume that log-normal distribution is true or precisely measured distribution and normal distribution is the approximated distribution of log-normal distribution. 
\begin{eqnarray} \nonumber
    &p(\bfx) = \mathcal{N}(\bfx|\bfm_p,\bfSigma_p) = c~ |\bfSigma_p|^{-1/2}\exp\Big[ -\frac{1}{2} ||\bfSigma_p^{-1/2}\big( \bfx -\bfm_p \big)||^2 \Big]\\ \nonumber
    &q(\bfx) = \log\mathcal{N}(\bfx|\bfm_q,\bfSigma_q) = c~ |\bfSigma_q|^{-1/2} \exp \Big[ -\frac{1}{2} ||\bfSigma_q^{-1/2}\big( \log (\bfx) -\bfm_q \big)||^2  \Big] \Pi^n_{i=1} x_i^{-1}
\end{eqnarray}
where $c=(2\pi)^{-n/2}$, and $\bfm_q$ and $\bfSigma_q$ are mean and covariance of $\log (\bfx)$, respectively. 
Consequently, the KL-divergence is defined as:
\begin{eqnarray*} 
    &D_\text{KL}(q(\bfx)||p(\bfx)) = \int q(\bfx) \log\frac{q(\bfx)}{p(\bfx)} = \E_q \big[ \log \big(q(\bfx)\big) - \log \big(p(\bfx)\big) \big]\\
    &\log\big(q(\bfx)\big) = -\frac{n}{2}\log(2\pi) -\frac{1}{2}\log |\bfSigma_q| -\frac{1}{2}||\bfSigma_q^{-1/2}\big( \log (\bfx) -\bfm_q \big)||^2 -\sum^n_{i=1} \log(x_i)\\
    &\log\big(p(\bfx)\big) = -\frac{n}{2}\log(2\pi) -\frac{1}{2}\log |\bfSigma_p| -\frac{1}{2}||\bfSigma_p^{-1/2}\big( \bfx -\bfm_p \big)||^2\\
    &\E_q \big[ \log\big(q(\bfx)\big) \big] = -\frac{n}{2}\log(2\pi) -\frac{1}{2}\log |\bfSigma_q| -\frac{1}{2}\trace (\bfI_{n\times n}) - \sum^n_{i=1} m_q^i\\
    &\E_q \big[ \log\big(p(\bfx)\big) \big] = -\frac{n}{2}\log(2\pi) -\frac{1}{2}\log |\bfSigma_p| - \frac{1}{2} \E_q\Big[ ||\bfSigma_p^{-1/2}\big( \bfx -\bfm_p \big)||^2 \Big]
\end{eqnarray*}
where instead of $(\bfx-\bfm_p)^\top \bfSigma_p^{-1}(\bfx-\bfm_p)$ we can write $\trace\{(\bfx-\bfm_p)^\top \bfSigma_p^{-1}(\bfx-\bfm_p)\}$, which can be re-written as $\trace\{(\bfx-\bfm_p)(\bfx-\bfm_p)^\top \bfSigma_p^{-1}\}$, and it can be shown that:
\begin{align} \nonumber
    \frac{1}{2} \E_q\Big[ ||\bfSigma_p^{-1/2}\big( \bfx -\bfm_p \big)||^2 \Big] =  \frac{1}{2} \E_q\Big[ \trace\Big\{(\bfx-\bfm_p)(\bfx-\bfm_p)^\top   \bfSigma_p^{-1}\Big\} \Big] \\ \nonumber
    = \frac{1}{2} \trace\Big\{\E_q\Big[ (\bfx-\bfm_p)(\bfx-\bfm_p)^\top \Big]  \bfSigma_p^{-1} \Big\}
\end{align}
where $\E_q\big[ (\bfx-\bfm_p)(\bfx-\bfm_p)^\top \big] = \Tilde{\bfSigma}_q + (\Tilde{\bfm}_q -\bfm_p) (\Tilde{\bfm}_q -\bfm_p)^\top$. Therefore, $ \E_q \big[ \log\big(p(\bfx)\big) \big]$ can be derived as:
\begin{equation} \nonumber
    \E_q \big[ \log\big(p(\bfx)\big) \big] = -\frac{n}{2}\log(2\pi) -\frac{1}{2}\log |\bfSigma_p| - \frac{1}{2} \trace\{ (\Tilde{\bfSigma}_q + (\Tilde{\bfm}_q -\bfm_p) (\Tilde{\bfm}_q -\bfm_p)^\top) \bfSigma_p^{-1} \}
\end{equation}
and the KL-divergence will be:
\begin{equation} \nonumber
    D_\text{KL}\big(q(\bfx)||p(\bfx)\big) =  -\frac{1}{2}\log \frac{|\bfSigma_q|}{|\bfSigma_p|} -\frac{n}{2} - \sum^n_{i=1} m_q^i + \frac{1}{2} \trace\Big\{ \big(\Tilde{\bfSigma_q} + (\Tilde{\bfm_q} -\bfm_p) (\Tilde{\bfm_q} -\bfmu_p)^\top\big) \bfSigma_p^{-1} \Big\}
\end{equation}
where $\Tilde{\bfSigma}_q$ and $\Tilde{\bfm}_q$ are covariance and mean of $\bfx$ in the log-normal distribution, which can be derived as follows:
\begin{eqnarray*}
    &\Tilde{\Sigma}_q^{ij} = \exp \big[ m_q^i + m_q^j + 0.5 \left(\Sigma_q^{ii} +\Sigma_q^{jj}\right) \big] \big( \exp (\Sigma_q^{ij}) -1 \big)\\
    &\Tilde{\bfm}_q = \exp\left( \bfm_q + 0.5 \bfSigma_q \bf1 \right)
\end{eqnarray*}

\section{Sampling from High-dimensional Log-normal Distribution}\label{sec:Appendix D}

In this section, we consider a high-dimensional log-normal distribution, the mean value and covariance of $\log(\bfpsi)$ is assumed as the covariance and mean value of the first example (section \ref{sec:Gaussian Posterior Distribution}). 
Since the range of $\bfpsi$ value is small, the negative and zero values for the log-normal distribution are not defined; taking samples from this distribution is complicated. To overcome this problem, we increase the mean value by adding a constant vector $\bfm' = \bfm + c \bf1 \Rightarrow \bfpsi' \sim \log \mathcal{N}(\bfm + c \bf1, \bfSigma)$.

We apply MH-MCMC, HMC, and H-HMC (with the Hessian at the MAP point) to explore the log-normal distribution. The step size is assumed $\Delta t = 1$ for MH-MCMC and HMC methods, and $\Delta t = 0.01$ for the H-HMC method, and the acceptance rate for all of these methods is approximately $98\%$.

We take $5\times10^5$ samples to explore the log-normal distribution; since the $\bfpsi'$ covers a wide range of values, the convergence speed of MH-MCMC and HMC methods is significantly lower than the convergence speed of the H-HMC method. Using the Hessian information as the mass matrix in the H-HMC method modifies the sampling process to explore the distribution in different directions (modes) faster. 
Even after $5\times10^5$ samples MH-MCMC, and HMC methods are far from convergence. However, H-HMC sampling chains converge after $2\times10^5$ samples. 



\end{document}